\documentclass[showpacs,aps,prd,nofootinbib,floatfix,amsmath,amssymb]{revtex4}
\usepackage{graphicx}
\usepackage[dvipsnames]{xcolor}
\usepackage{dsfont}
\usepackage{slashed}
\begin{document}

\makeatletter
\newbox\slashbox \setbox\slashbox=\hbox{$/$}
\newbox\Slashbox \setbox\Slashbox=\hbox{\large$/$}
\def\pFMslash#1{\setbox\@tempboxa=\hbox{$#1$}
  \@tempdima=0.5\wd\slashbox \advance\@tempdima 0.5\wd\@tempboxa
  \copy\slashbox \kern-\@tempdima \box\@tempboxa}
\def\pFMSlash#1{\setbox\@tempboxa=\hbox{$#1$}
  \@tempdima=0.5\wd\Slashbox \advance\@tempdima 0.5\wd\@tempboxa
  \copy\Slashbox \kern-\@tempdima \box\@tempboxa}
\def\FMslash{\protect\pFMslash}
\def\FMSlash{\protect\pFMSlash}
\def\miss#1{\ifmmode{/\mkern-11mu #1}\else{${/\mkern-11mu #1}$}\fi}
\makeatother

\title{About heavy neutrinos: Lepton-flavor violation in decays of charged leptons}

\author{H. Novales-S\'anchez}
\author{M. Salinas}
\author{J. J. Toscano}
\affiliation{Facultad de Ciencias F\'{\i}sico Matem\'aticas,
Benem\'erita Universidad Aut\'onoma de Puebla, Apartado Postal
1152, Puebla, Puebla, M\'exico.}

\begin{abstract}
The fundamental description of nature, beyond the Standard Model (SM), may include heavy neutrinos that mix and thus allow processes in which lepton flavor is not preserved. We investigate the impact of charged currents that couple heavy gauge bosons to heavy neutrinos and SM leptons on lepton-flavor-violating decays of SM leptons into three charged leptons, with no final-state neutrinos. We implement our expressions for the leading contributions to ${\rm Br}(l_\alpha\to l_\beta\,l_\sigma\,l_\sigma)$, which hold for either Dirac or Majorana neutrinos, to the trilepton decay $\mu\to3e$, of the muon, and so determine sets of masses of heavy neutrinos and the heavy gauge boson, within GeVs to few TeVs, that are consistent with the upper bounds provided by the SINDRUM Collaboration. 
We find, however, that constraints dictated by the upper bound on ${\rm Br}(\mu\to e\gamma)$, from the MEG Collaboration, are more stringent.
We utilize such parameters to find that the contributions to tau decays are $\sim10^{-15}-10^{-13}$, well below bounds from $B$ factories. 
The mixing of heavy and SM charged bosons is also investigated. We find that current experimental data from MEG and SINDRUM would allow mixing angles as large as $\sim10^{-2}$, for a relatively light new charged boson, but the expected sensitivity of the Mu3e experiment would be capable of setting an upper bound on this angle as small as $\sim10^{-4}$ if the mass of this boson is within the range of few TeVs.
\end{abstract}

\pacs{11.30.Hv, 13.35.-r, 14.60.St, 14.70.Pw}

\maketitle
\section{Introduction}
\label{int}
Neutrino physics is nowadays a topic of interest that has experienced great progress. This includes the determination of the whole set of neutrino-mixing angles~\cite{PDG,EGMMS} from data acquired in experiments that access the phenomenon of neutrino oscillations, which is clear experimental evidence that neutrinos are massive and mix~\cite{Pontecorvo,GK}, thus indicating the presence of new physics beyond the Standard Model. Neutrino oscillations were first observed at the Super-Kamiokande~\cite{SKKnus} and then confirmed at the Sudbury Neutrino Observatory~\cite{SNOnus}. The last mixing angle, $\theta_{13}$, was finally measured, almost simultaneously, by the Daya Bay and RENO Collaborations~\cite{Dayanus,Renonus}, though it must be pointed out that just months before this the Double Chooz experiment announced, for the first time, that $\theta_{13}\ne0$~\cite{DChooz}. Experimental investigations intended to determine the Dirac $CP$-violating phase, such as that performed by the T2K Collaboration in Ref.~\cite{T2K}, also exist. One of the main unanswered questions in neutrino physics concerns whether neutrinos are described by Dirac fields, as the rest of the Standard-Model fermions, or they correspond to Majorana particles\footnote{Interestingly, recent experimental evidence, in solid-state physics, of the existence of a Majorana fermion state, an analogue of the Majorana states not yet observed in experiments of high-energy physics, was found~\cite{MajFer}.}, as first realized by Ettore Majorana~\cite{Majorana}. Experimental setups aiming at the observation of the elusive {\it neutrinoless double-beta decay} have been based on the double beta decay of diverse isotopes: $^{76}$Ge (Heidelberg-Moscow, IGEX, GERDA)~\cite{HeidelbergMoscow,IGEX,GERDA}, $^{100}$Mo and $^{82}$Se (NEMO)~\cite{NEMO1,NEMO2}, $^{130}$Te (CUORICINO)~\cite{CUORICINO}, $^{150}$Nd (NEMO)~\cite{NEMO3}, and $^{136}$Xe (KamLAND-Zen, EXO)~\cite{KamLandZen1,EXO,EXO200}. While a measurement of this decay would be evidence in favor of Majorana neutrinos, so far this process has not been observed in nature~\cite{KamLandZen2}. Interestingly, electromagnetic properties of Dirac neutrinos are quite different from those characterizing Majorana fermions~\cite{SchVa,Kayser,Shrock,Nieves,emnus1,emnus2}. The fact that neutrinos, even being electrically neutral, can interact with the electromagnetic field through quantum effects has been investigated in a variety of works, which have explored the neutrino anapole moment and charge radius~\cite{RMandH,BPV,FuSh,NRST}, the neutrino electric dipole moment~\cite{GrNe,ADD,DvSt} and, most frequently, neutrino magnetic moments~\cite{Kim,BMR,CGZ,MoPl,DvSt2,BGRVW,BoBo,DeNo}. There is even the proposal that neutrinos might not have zero electric charge, but they could be millicharged particles instead~\footnote{See Ref.~\cite{GiSt} and references therein.}. \\

The {\it seesaw mechanism}~\cite{MoSeseesaw}, included by field theories with the purpose of explaining neutrino mass, show us that the high-energy description of fundamental physics, beyond the Standard Model, may include heavy neutrinos with masses depending on some energy scale, $\Lambda$, characterizing the formulation. 
Heavy neutrinos might also arise as Kaluza-Klein excited modes that originate in theories set on extra-dimensional spacetime manifolds~\cite{DDG,ADPY,BMOZ}; the Kaluza-Klein heavy-neutrino masses are determined by some high-energy compactification scale $\Lambda$, which is defined by the size of the extra dimensions. 
With all this in mind, we consider the charged currents
\begin{equation}
\sum_{j=1}^3\sum_{\alpha=e,\mu,\tau}\left[ \frac{v_{j\alpha}}{\sqrt{2}}\,W'^+_\rho\bar{N}_j\gamma^\rho P_Rl_\alpha+{\rm H.c.} \right],
\label{CCs}
\end{equation}
where $N_j$ has been used to denote three heavy neutrinos that couple to a heavy charged gauge boson $W'$ and to Standard-Model charged leptons $l_\alpha$. 
We have used the coefficients $v_{j\alpha}$ to characterize the mixing of heavy neutrinos in these charged currents, which is an essential ingredient in the present investigation as it allows lepton-flavor-violating processes that are forbidden in the context of the Standard Model. We shall assume that these coefficients are the entries of a $3\times3$ matrix that is approximately unitary\footnote{By ``approximately unitary'' we mean that $\sum_{j=1}^3v^*_{j\alpha}v_{j\alpha}\approx1$, while for $\alpha\ne\beta$ the condition $\sum_{j=1}^3v^*_{j\beta}v_{j\alpha}=0$ holds.}.
We assume that $\Lambda$ is large and that, under such circumstances, both the heavy-neutrino masses, $m_j$,  and the $W'$ mass, $m_{W'}$, are approximately proportional to this high-energy scale. Instances of formulations in which this occurs are {\it left-right symmetric models}~\cite{MoSeseesaw,MoPa,MoPa2,SeMo}, the {\it simplest little Higgs model}~\cite{dAIJ},  {\it 331 models}~\cite{CaMa}, and the {\it 5-dimensional extension of the Standard Model}~\cite{CGNT}. Thus, we note that $m_{W'}\approx\kappa_jm_{j}$, with $\kappa_j$ independent of $\Lambda$ for any $j$.
Charged currents like those given in Eq.~(\ref{CCs}) were considered before, in the context of the Large Hadron Collider, in Ref.~\cite{HLRZ}, where the authors analyzed the lepton-number-violating process $pp\to W'\to l^{\pm}N\to l^{\pm}l^{\pm}jj$, allowed in the presence of Majorana neutrinos~\cite{KeSe}. Right-handed currents similar to those of Eq.~(\ref{CCs}) were used in another work to carry out an investigation on the process $pp\to lljj$~\cite{GJS}.
\\

In the Standard Model, lepton-flavor violation is strictly forbidden and, as a consequence, the decays of charged leptons into three leptons are always characterized by final states that involve only one charged particle and two neutrinos, whereas decays into three charged leptons, with no neutrinos, are absent. Nevertheless, neutrino mixing admits of lepton-flavor-violating Feynman diagrams that contribute, since the one-loop level, to the latter type of charged-lepton decays. In such a context, the muon is able to decay like $\mu^-\to e^-e^+e^-$. Among the Standard Model leptons, the tau is the heaviest and thus the one with the richest decay spectrum~\cite{PDG,Pich}. Its large mass makes it possible for this particle to include the decays $\tau^-\to e^-\mu^+\mu^-$, $\tau^-\to\mu^-e^+e^-$, $\tau^-\to e^-e^+e^-$, and $\tau^-\to\mu^-\mu^+\mu^-$. In the present paper, we explore and analyze this possibility. We point out that, driven by neutrino mixing, the charged currents of Eq.~(\ref{CCs}) generate contributions from the heavy neutrinos $N_j$ and the heavy gauge boson $W'$ to these trilepton decays of the muon and the tau. Keeping things as model-independent as possible, we calculate leading one-loop contributions to lepton-flavor-violating decays $l_\alpha\to l_\beta\, l_\sigma\, l_\sigma$ and then implement our result to the aforementioned processes. We analyze such decays in detail in a scenario in which two heavy neutrinos have masses that are quasi-degenerate, but the mass of the third neutrino is different, with the masses $m_j$ and $m_{W'}$ within the range from hundreds of GeVs to few TeVs. \\

We find that, for given values of the kappa factors $\kappa_j$ that we defined before, a lower limit on the $W'$ mass $m_{W'}$ can be set so that the leading contributions from the charged currents of Eq.~(\ref{CCs}) to the branching ratio ${\rm Br}(\mu\to3e)$ are compatible with the upper bound provided by the SINDRUM Collaboration~\cite{SINDRUM}. 
Moreover, we also determine bounds on $m_{W'}$ through the branching ratio ${\rm Br}(\mu\to e\gamma)$, which has been constrained by the MEG Collaboration~\cite{MEGcoll}, and find that lower limits extracted from $\mu\to e\gamma$ are more stringent than those established by current limits on ${\rm Br}(\mu\to3e)$.
Motivated by the possibility of revealing the presence of new physics through the measurement of the muon decay process $\mu\to3e$~\cite{Mu3ecoll}, the {\it Mu3e experiment} is expected to observe more than $10^{16}$ muon decays. We determine that a level of sensitivity like that shall increase lower bounds on the $W'$ mass by one order of magnitude with respect to bounds obtained from SINDRUM. A discussion on the neutrino-mixing dependence of our results is carried out as well. We observe that, for given values of the parameters $\kappa_j$, the lower limits on the mass $m_{W'}$, established from the MEG bound on ${\rm Br}(\mu\to e\gamma)$, produce contributions to the afore-alluded tau decays of order $10^{-15}-10^{-13}$, which are in accordance with the bounds from $B$ factories~\cite{BABARLFVtau,BelleLFVtau}. We find it worth emphasizing that the diagrams generating the dominant contributions are the same no matter whether the heavy neutrinos are Dirac or Majorana fermions, so our conclusions practically hold in both cases. \\

A physical situation in which the presumed heavy charged boson $W'$ mixes with the Standard-Model gauge boson $W$ is plausible~\cite{SenMoother}, and if this is the case, then the parameter characterizing the $W'$-$W$ mixing is a {\it mixing angle}, $\zeta$. The effects of such a mixing might impact charged currents in which heavy neutrinos are coupled with the $W$ boson and the charged leptons of the Standard Model~\cite{HLRZ}. We calculate the contributions from these mixing effects to both ${\rm Br}(\mu\to e\gamma)$ and ${\rm Br}(\mu\to3e)$, and then use the bounds from MEG and SINDRUM to establish allowed regions in the parameter space $(|\zeta|,m_{W'})$. We point out that, while these bounds can be compatible with a relatively large value of the mixing angle if $m_{W'}$ is small, larger values of the $W'$ mass push this bound towards smaller angles $\zeta$. Furthermore, we estimate that the projected sensitivity of Mu3e is able to improve upper limits on $|\zeta|$ by even two orders of magnitude. \\

We have organized this document in the following manner: in Section \ref{calc}, we calculate the leading contributions to the process $l_\alpha\to l_\beta\, l_\sigma\, l_\sigma$ and show that our expressions are ultraviolet finite and decoupling; then we specialize to $l_\alpha=\mu^-,\tau^-$, in Section~\ref{DandR}, and consider the decays $\mu^-\to e^-e^+e^-$, $\tau^-\to e^-e^+e^-$, $\tau^-\to \mu^-\mu^+\mu^-$, $\tau^-\to e^-\mu^+\mu^-$ and $\tau^-\to\mu^-e^+e^-$, whose branching ratios are calculated and analyzed, for which the decay $\mu\to e\gamma$ is taken into account as well; in Section~\ref{secmix},  the decay process $\mu^-\to e^-e^+e^-$ is considered again in order to discuss $W'$-$W$ mixing, and bounds on the corresponding mixing angle are derived; we end the paper by presenting our conclusions in Section~\ref{conc}.\\


\section{The trileptonic decay $l_\alpha\to l_\beta\, l_\sigma\, l_\sigma$}
\label{calc}
In this section, we calculate one-loop contributions from the charged currents given in Eq.~(\ref{CCs}) to the invariant matrix element ${\cal M}_{\alpha\to\beta\sigma\sigma}$, of the lepton-flavor-violating decay $l_\alpha\to l_\beta\,l_\sigma\, l_\sigma$.

\subsection{The amplitude ${\cal M}_{\alpha\to\beta\sigma\sigma}$}
\begin{figure}[!ht]
\center
\includegraphics[width=5.3cm]{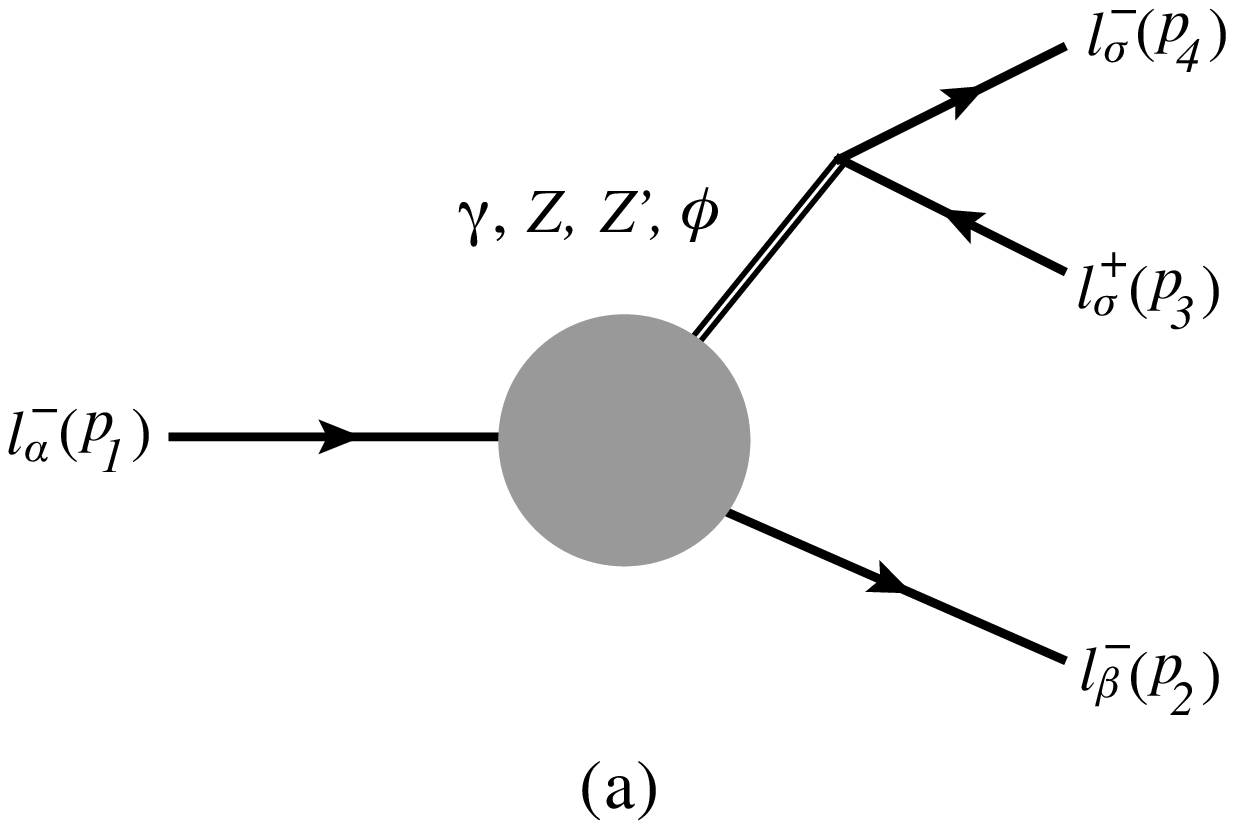}
\hspace{0.6cm}
\includegraphics[width=5.3cm]{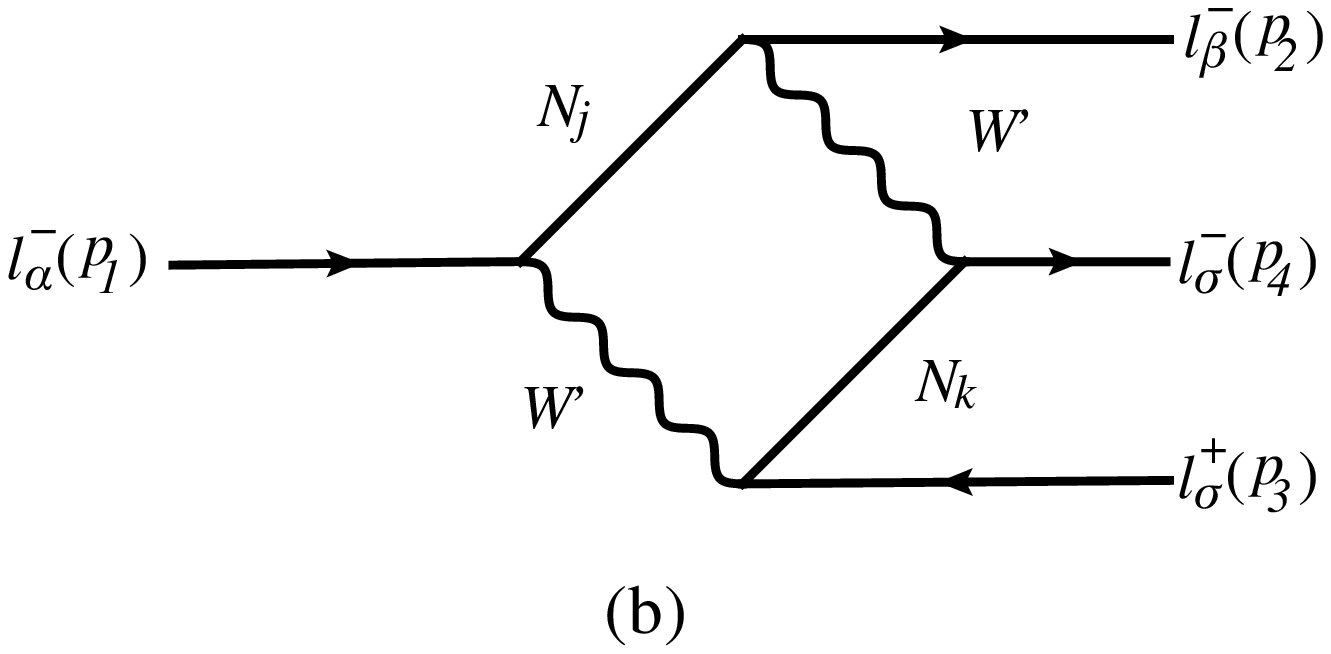}
\hspace{0.6cm}
\includegraphics[width=5.3cm]{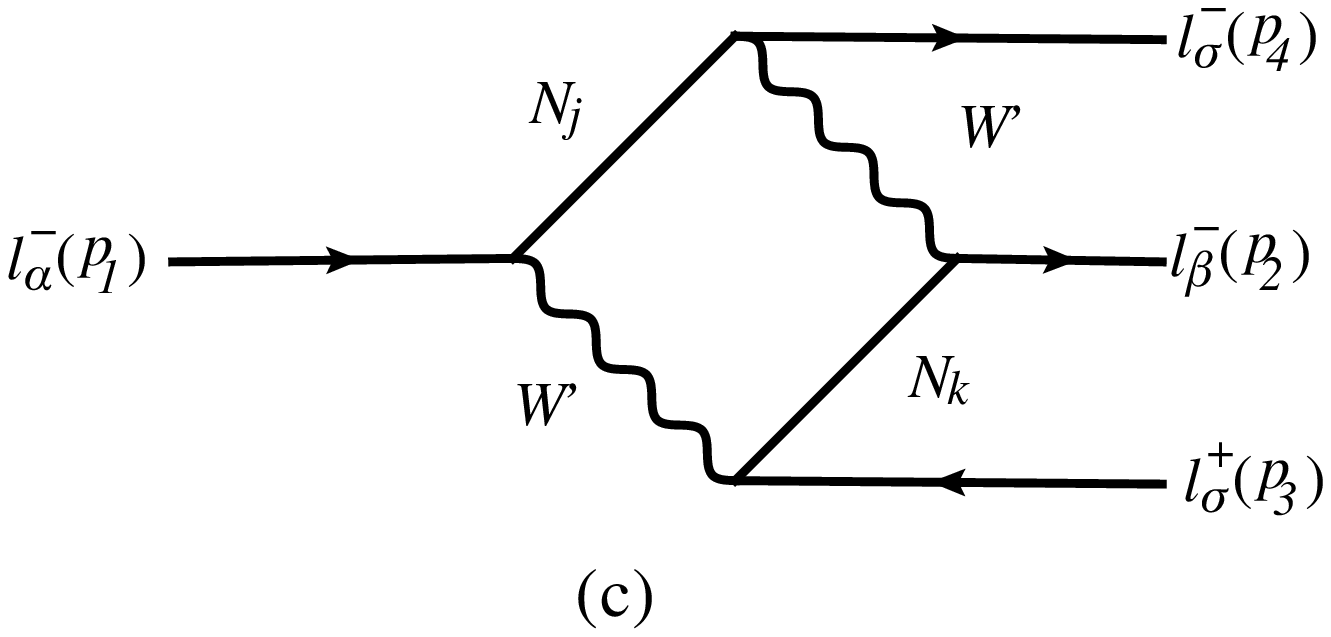}
\caption{\label{decdiags} Feynman diagrams that produce contributions, at one loop, to the decay $l_\alpha\to l_\beta\, l_\sigma\, l_\sigma$ by means of neutrino mixing: (a) reducible diagrams; and (b) and (c) box diagrams.}
\end{figure}

At one loop, the charged currents given in Eq.~(\ref{CCs}) produce two types of Feynman diagrams that contribute to $l_\alpha\to l_\beta\,l_\sigma\,l_\sigma$: (a) reducible diagrams, with three-leg loop subdiagrams; (b) and (c) box diagrams. We show both sorts of diagrams in Fig.~\ref{decdiags}~\footnote{Diagrams (b) and (c), shown in this figure, are the only box diagrams if the neutrinos $N_j$ are Dirac fermions. Neverthelss, in the case of Majorana neutrinos there are additional box diagrams, since in such a context two charged leptons with the same sign can be emitted from a neutrino line.}.
In this figure, the virtual-particle double line in the generic reducible diagram of type (a) represents a propagator that can correspond, in general, to a photon, a Standard Model $Z$ boson, a heavy $Z'$ boson, or some neutral scalar $\phi$. For each of these propagators there is a set of one-loop irreducible subdiagrams, which we have represented by the shadowed circle. \\

The structure of any diagram of Fig.~\ref{decdiags} bears gauge dependence, although the sum of all the contributing diagrams must be independent of the chosen gauge. It is a common practice to use the so-called {\it unitary gauge} in order to study new unknown physics. The absence of unphysical pseudo-Goldstone bosons, with the corresponding decrease of the number of contributing diagrams, is a practical advantage of the unitary gauge. However, in this gauge the cancellation of ultraviolet divergences is nontrivial and requires very specific information about the model behind the charged currents of Eq.~(\ref{CCs}). For instance, think for a moment in the the addition of three right-handed Dirac-neutrino singlets to the Standard Model, which yields a {\it Minimally Extended Standard Model} that includes three light Dirac neutrinos with masses generated by the Englert-Higgs mechanism~\cite{Higgs1,Higgs2,EngBr}, with the sizes of such masses determined by Yukawa constants and by a vacuum expectation value~\cite{GK}. Within such a context, mixing of light neutrinos arises, and the flavor-changing decay $l_\alpha\to l_\beta l_\sigma l_\sigma$ is thus allowed. It turns out that the sum of the subdiagrams $l_\alpha l_\beta\gamma^*$ ($\gamma^*$ denotes a virtual photon) that are part of the reducible diagrams of Fig.~\ref{decdiags} (with light neutrinos and the Standard-Model $W$ boson replacing the corresponding heavy particles in this figure) is finite if they are calculated on shell~\cite{NSTV}, $\gamma^*\to\gamma$. But if the photon is off shell instead this sum is ultraviolet divergent. The sum of all the irreducible off-shell subdiagrams $l_\alpha l_\beta Z^*$ is also ultraviolet divergent. A combination of all the corresponding reducible diagrams (both with virtual $\gamma$ and virtual $Z$) and the box diagram (b), in Fig.~\ref{decdiags}, is required for an intricate cancellation to occur, and thus to obtain finite expressions. The box diagram (c), on the other hand, is ultraviolet finite by itself. The delicate balance that produces such a cancellation depends crucially on the specific couplings of the $Z$ boson to neutrinos and charged leptons; a rho parameter $\rho=1$, meaning that $m_W=m_Z\cos\theta_W$, is necessary as well. \\

A useful parametrization of a set of different gauges is embodied by the well-known {\it $R_\xi$ gauge}~\cite{FLS}. The $R_\xi$ gauge is characterized by a gauge-fixing parameter, $\xi$, which is a real number whose different values correspond to different gauges. As it must be in any gauge other than the unitary gauge, calculations in the $R_\xi$ gauge involving virtual massive gauge fields include contributions from pseudo-Goldstone bosons. Nevertheless, there are other gauges that may render the calculation simpler. Particularly, we found it convenient to perform the calculation in the {\it nonlinear $R_\xi$ gauge} that is widely discussed in Ref.~\cite{MenTos}, in the context of the Becchi-Rouet-Stora-Tyutin quantization~\cite{BRS1,BRS2,Tyutin,GPS}. Nonlinear gauges have been useful in phenomenology of, for instance, 331 gauge models~\cite{TaTo,MTTR} and Kaluza-Klein theories~\cite{NoTo,CGNT,LMMNTT,GNT}. In our case, the nonlinear gauge would be characterized by the gauge-fixing functions
\begin{equation}
f^{\pm}=D_\mu W'^{\pm\mu}-i\xi\, m_{W'} G'^{\pm}_{W'},
\end{equation}
where $D_\mu$ denotes the U(1)$_e$ covariant derivative, so that
this gauge-fixing scheme is covariant with respect to the electromagnetic group. Here, we have used $G_{W'}$ to denote the pseudo-Goldstone boson associated to $W'$. The use of this nonlinear gauge would eliminate unphysical trilinear couplings $W'G_{W'}\gamma$, which mix pseudo-Goldstone bosons $G_{W'}$ and gauge bosons $W'$, thus reducing the number of contributing Feynman diagrams. It is worth commenting, however, that a U(1)$_e$-covariant nonlinear $R_\xi$ gauge would not necessarily remove unphysical mixings $W'G_{W'}Z$ nor $W'G_{W'}Z'$ from the theory\footnote{In the context of the Standard Model, the nonlinear gauge fixing eliminates both the $WG_W\gamma$ and the $WG_WZ$ couplings~\cite{MenTos}.}. Therefore, the corresponding vertices would indeed be pieces of reducible diagrams of Fig.~\ref{decdiags}, with either $Z$ or $Z'$ loop-connecting propagator. As we explain below, these diagrams do not produce leading contributions and thus will be neglected.\\

The afore-described complicated cancellation of ultraviolet divergences that takes place in the unitary gauge merges the contributions from the participating diagrams in a nontrivial way. As a consequence, the finite contributions from each individual diagram cannot be discerned from the total contribution anymore. On the other hand, in the nonlinear gauge the cancellation of ultraviolet divergences is much simpler: the sum of all the irreducible subdiagrams $l_\alpha l_\beta \gamma^*$ is finite by itself \footnote{The same holds, in the context of the Standard Model with three massive Dirac neutrinos, for the sum of irreducible subdiagrams $l_\alpha l_\beta Z^*$, which happens independently of the couplings of $Z$ to neutrinos and charged leptons, and independently of the rho parameter.}. Moreover, in this gauge both box diagrams turn out to be finite as well. Then in the nonlinear gauge the individual contributions from the diagrams are clearly separated. In this context, it makes sense to assume that the contributions from reducible diagrams dominate over those generated by box diagrams, so from here on we neglect the contributions from the latter diagrams. \\

Concerning the reducible diagrams, the virtual line that connects irreducible loop subdiagrams with the external fermion current introduce a factor $1/q^2$, in the case of a virtual photon, and $1/(q^2-m^2)$ when this propagator corresponds to a massive boson of mass $m$. Our goal is to calculate the decay rate for $l_\alpha\to l_\beta l_\sigma l_\sigma$, which, as we show later, restricts $q^2$ to very small values: $4m_\sigma^2<q^2<(m_\alpha-m_\beta )^2$. In this context, the squared mass $m^2$ of any massive virtual particle that can contribute is very large with respect to $q^2$, so the factor $1/(q^2-m^2)$ suppresses all these contributions, whereas the factor $1/q^2$, in the case of the virtual photon, rather enhances the corresponding contribution. A similar situation occured in Ref.~\cite{CHTT}, where the flavor-changing trilinear quark decays $t\to u_1 \bar{u}_2u_2$, of the top quark into up-type quarks, were investigated at the one-loop level, in the context of the Standard Model. Taking this discussion into account, we calculate only the contributions from the type-(a) diagrams with the virtual photon and neglect all other contributions in what follows. \\

\begin{figure}[!ht]
\center
\includegraphics[width=5cm]{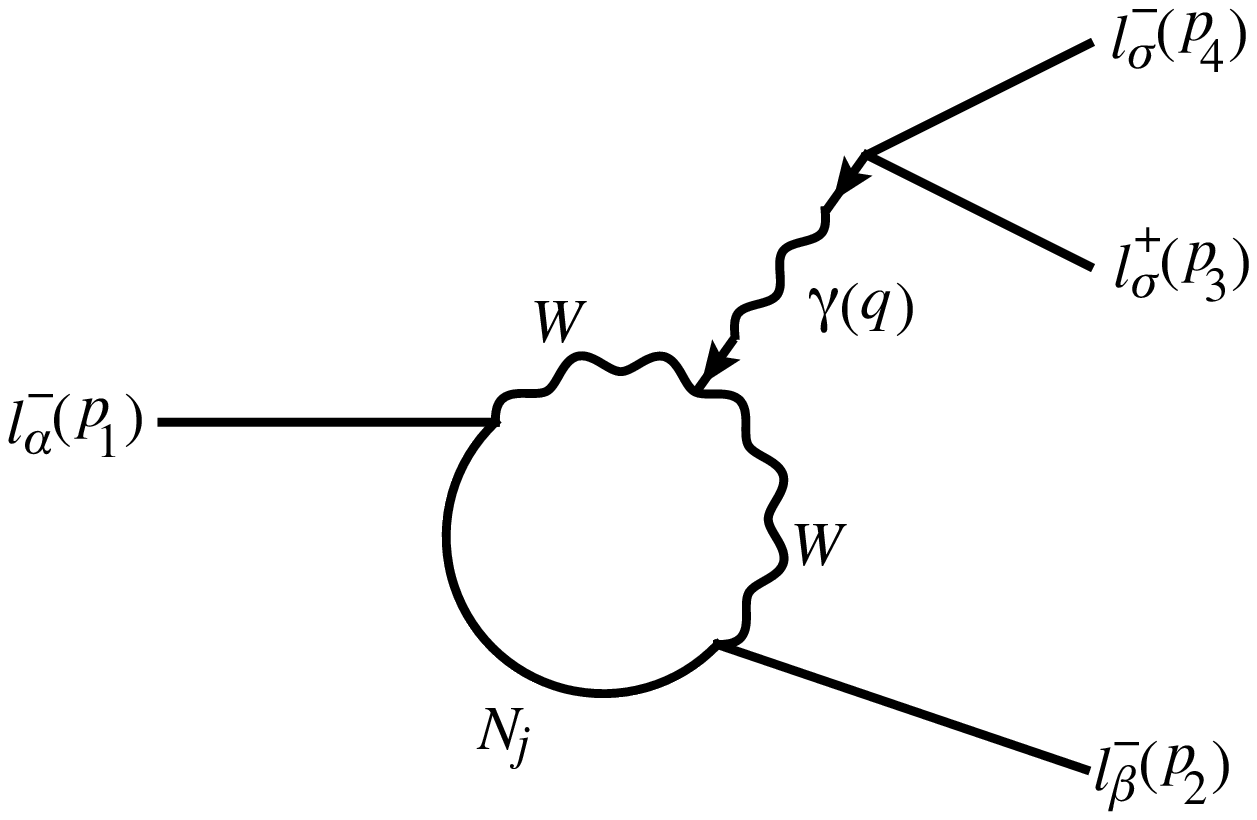}
\vspace{0.6cm}
\includegraphics[width=5cm]{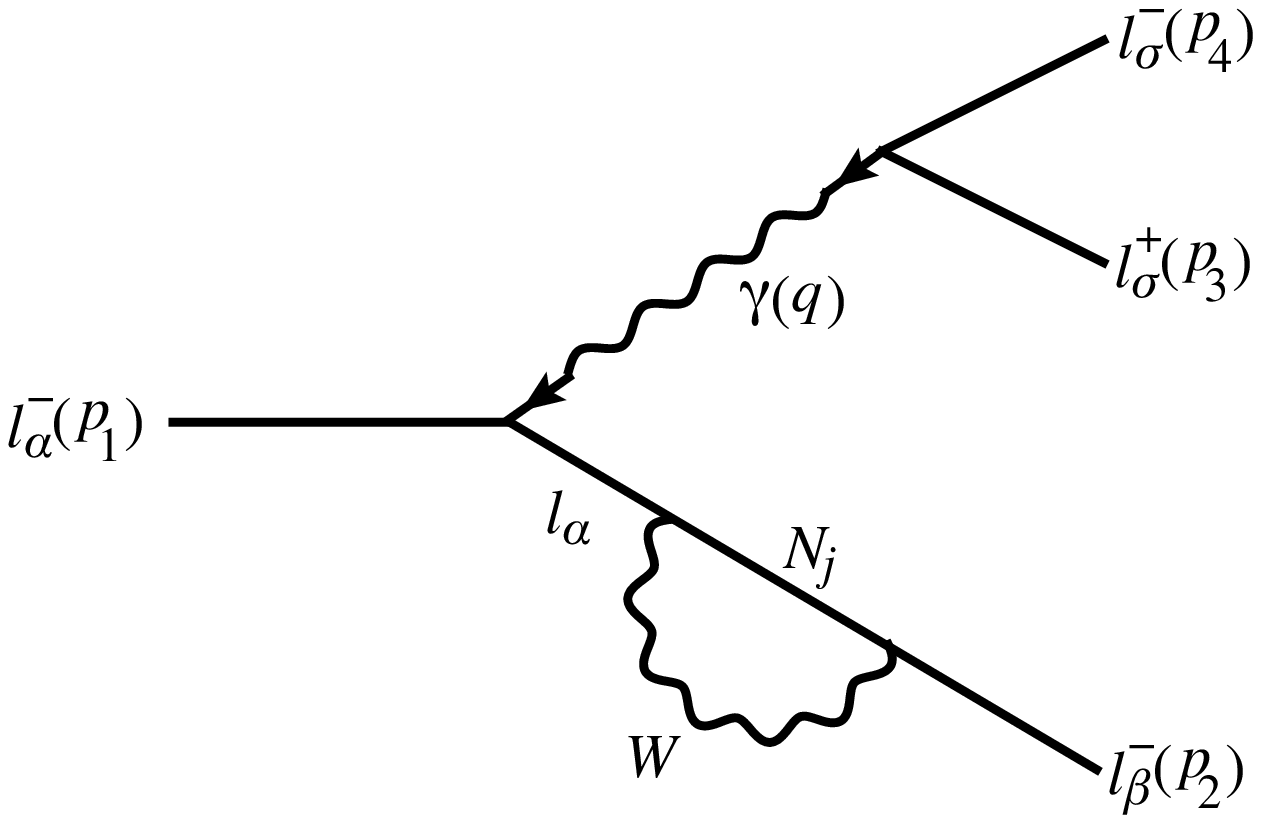}
\vspace{0.6cm}
\includegraphics[width=5cm]{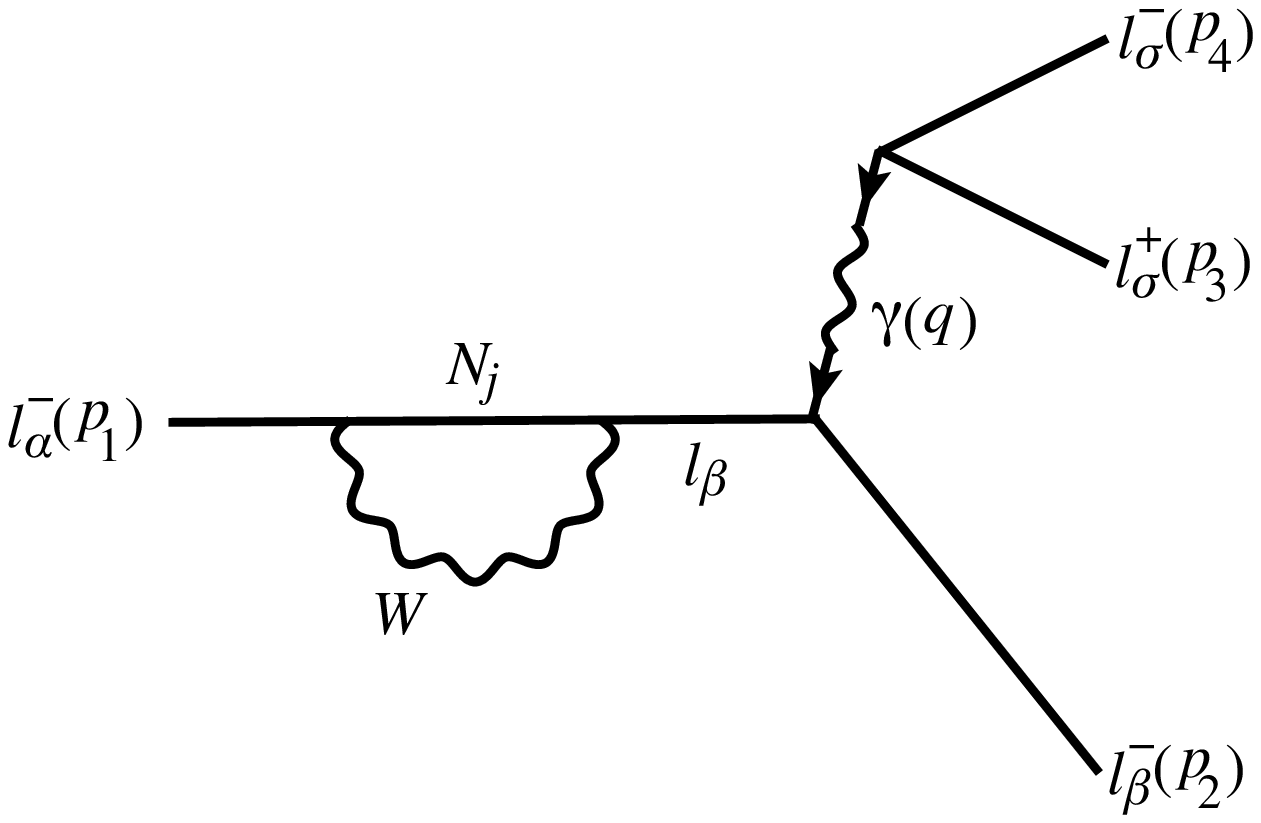}
\caption{\label{emdiagrams} Feynman diagrams corresponding to the dominant contribution from the charged currents of Eq.~(\ref{CCs}) to the decay $l_\alpha\to l_\beta\, l_\sigma\, l_\sigma$.}
\end{figure}

The leading contribution from the charged currents of Eq.~(\ref{CCs}) to the amplitude of $l_\alpha\to l_\beta\,l_\sigma\,l_\sigma\,$ is produced by the Feynman diagrams shown in Fig.~\ref{emdiagrams}.
Besides these, there are contributing diagrams that involve pseudo-Goldstone bosons, which must be taken into account. In the nonlinear $R_\xi$ gauge that we use, the diagrams with pseudo-Goldstone bosons $G_{W'}$ look exactly the same as those provided in Fig.~\ref{emdiagrams}, but with the $W'$-boson lines replaced by $G_{W'}$ lines. In this nonlinear gauge, both gauge-boson propagators and gauge vertices depend on the gauge-fixing parameter $\xi$. In particular, the coupling $W'W'\gamma$ yields the Feynman rule~\cite{MenTos}
\[
\raisebox{-11mm}{\includegraphics[width=4cm] {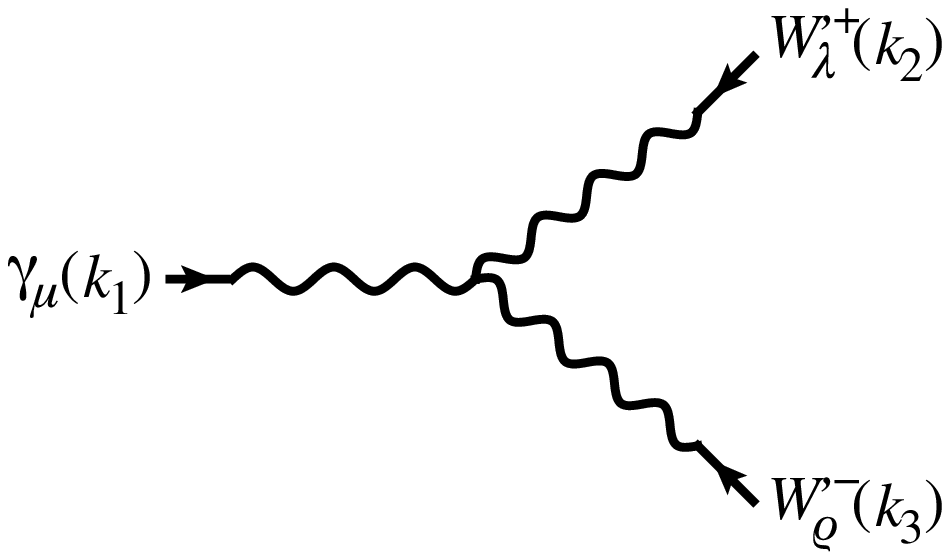}}
\begin {aligned}
=ie\Big[ (k_{3\eta}-k_{2\eta}) g_{\rho\lambda}
+\Big(k_{1\lambda}
-k_{3\lambda}-\frac{1}{\xi}k_{2\lambda}\Big) g_{\rho\eta}
+\Big(k_{2\rho}
-k_{1\rho}+\frac{1}{\xi}k_{3\rho}\Big) g_{\lambda\eta} \Big],
\end {aligned}
\]
which we set in a {\it Feynman-'t Hooft}-like gauge, that is, we take $\xi\to1$ to perform the calculation. About the unphysical vertex $G_{W'}G_{W'}\gamma$, required for this calculation, we assume that it is just like the one that characterizes the Standard Model~\cite{ChengLi}. 
\\

Neutrino oscillations~\cite{Pontecorvo} is a quantum phenomenon that consists in the occurrence of nonzero transition probabilities of measuring a neutrino flavor that is different from the neutrino flavor originally emited at certain source. The observation of neutrino oscillations has been meaningful, in part because it incarnates solid evidence that neutrinos have nonzero mass. Furthermore, neutrinos, being electrically neutral and massive, can be described by either Dirac or Majorana fields~\cite{GK,Majorana}. 
Dirac neutrinos differ from Majorana ones in several matters. In particular, the Feynman rules for these cases are different of each other. The Feynman rules for Majorana fermions have been already derived and discussed in detail in Refs.~\cite{DEHK,GluZra}. From Wick's theorem~\cite{Wick,PeSch}, it can be seen that, in general, the number of Feynman diagrams to consider in the Majorana case is different from that of Dirac fermions. For instance, if the neutrinos $N_j$ couple to the $Z$ boson and we wish to calculate the one-loop contribution to the off-shell vertex $l_\alpha l_\beta Z^*$, the Majorana case involves one additional contributing diagram with respect to the Dirac case. However, in particular, the calculation of the contributions from reducible diagrams with the one-loop electromagnetic vertex, which dominates, turns out to be the same in both cases. This means that, in a practical sense, our results are valid no matter whether the heavy neutrinos are Dirac or Majorana fermions. \\

We have utilized the method of {\it Passarino-Veltman reduction}~\cite{PaVe}, which leaded us to expressions given in terms of scalar functions $B_0$ and $C_0$. We have also used the software {\it Mathematica}, by Wolfram, with the package {\it Feyncalc}~\cite{SMO,MBD}. According to our previous discussion, there are six one-loop three-line irreducible electromagnetic subdiagrams involved in the dominant contribution to the amplitude of $l_\alpha\to l_\beta \,l_\sigma \,l_\sigma$. The sum of such subdiagrams produces the total contribution $\bar{u}_\beta(p_2)\Gamma^{\alpha\beta}_\mu(q)u_\alpha(p_1)$, with the general gauge-invariant structure\footnote{By ``gauge-invariant structure'' we mean that gauge invariance holds with respect to the electromagnetic gauge group, and thus the Ward identity $q^\mu\Gamma_\mu^{\alpha\beta}=0$ is fulfilled.}~\cite{emnus2,HIRSS,NPR}
\begin{eqnarray}
\Gamma_\mu^{\alpha\beta}(q)&=&ie\sum_{j=1}^3 v_{j\beta}^*v_{j\alpha}\Big[\Big( \gamma_\mu-\frac{q_\mu \slashed{q}}{q^2} \Big)\big( f^j_Q+f^j_Aq^2\gamma_5 \big)
-i\sigma_{\mu\nu}q^\nu\big(f^j_M+if^j_E\gamma_5\big)
\Big],
\label{emvpar}
\end{eqnarray}
which includes a sum over the heavy-neutrinos contributions, labeled by $j=1,2,3$. Any heavy-neutrino contribution in this expression is written in terms of the charge form factor $f^j_Q$, the anapole form factor $f^j_A$, the magnetic form factor $f^j_M$, and the electric form factor $f^j_E$. For a given $j$, these quantities depend on the masses $m_j$ and $m_{W'}$, on the masses $m_\alpha$ and $m_\beta$, corresponding to the leptons $l^-_\alpha$ and $l^-_\beta$, and on the squared four-momentum $q^2$ (see notation and conventions in Fig.~\ref{emdiagrams}). We write these contributions to electromagnetic form factors in the generic form
\begin{eqnarray}
f^j_\Omega&=&\frac{1}{\Delta_\Omega}\Big[ f^j_{\Omega,0}B^j_0+f^j_{\Omega,12}(B^j_1-B^j_2)+f^j_{\Omega,23}(B^j_2-B^j_3) 
+f^j_{\Omega,34}(B^j_3-B^j_4)+f^j_{\Omega,5}C^j_0 \Big],
\label{finfs}
\end{eqnarray}
where $\Omega=Q,A,M,E$. The explicit expressions of the factors $f^j_{\Omega,0}$, $f^j_{\Omega,12}$, $f^j_{\Omega,23}$, $f^j_{\Omega,34}$, and $f^j_{\Omega,5}$ are provided in Appendix~\ref{appfs}. Furthermore, we have used the following notation for the scalar functions that appear in Eq.~(\ref{finfs}):
\begin{equation}
B^j_1=B_0(0,m_{W'}^2,m_j^2),
\label{defB1}
\end{equation}
\begin{equation}
B^j_2=B_0(q^2,m_{W'}^2,m_{W'}^2),
\end{equation}
\begin{equation}
B^j_3=B_0(m^2_\alpha,m_{W'}^2,m_j^2),
\end{equation}
\begin{equation} 
B^j_4=B_0(m^2_\beta,m_{W'}^2,m_j^2),
\end{equation}
\begin{equation}
C^j_0=C_0(m^2_\alpha,m^2_\beta,q^2,m_{W'}^2,m_j^2,m_{W'}^2),
\label{defC0}
\end{equation}
and we defined $B^j_0=1$. Using the dimensional-regularization approach~\cite{BoGi}, any scalar function $B_0$ can be written in the form $B_0=\Delta_{\rm div}+f_{\rm fin}$~\cite{tHV}, with all the ultraviolet divergences and the logarithmic cutoff dependence contained in $\Delta_{\rm div}$. From this general expression of any $B_0$ the main feature of Eq.~(\ref{finfs}) is clear: it shows in an explicit manner that each contribution $f^j_\Omega$ is, by itself, finite in the ultraviolet sense, meaning that the whole contribution $\Gamma_\mu^{\alpha\beta}$ is free of ultraviolet divergences. \\

The amplitude ${\cal M}_{\alpha\to\beta\sigma\sigma}$, for the process $l_\alpha\to l_\beta\, l_\sigma\, l_\sigma$, is given by
\begin{eqnarray}
i{\cal M}_{\alpha\to\beta\sigma\sigma}&=&\bar{u}_\beta(p_2)\Gamma_\mu^{\alpha\beta}u_\alpha(p_1)\frac{-ig^{\mu\nu}}{q^2}
\bar{u}_\sigma(p_4)(-ie\gamma_\nu)v_\sigma(p_3).
\label{invamp}
\end{eqnarray}
Due to the kinematics of the diagrams under consideration, it is possible to write ${\cal M}_{\alpha\to\beta\sigma\sigma}$ only in terms of the masses $m_j$, $m_{W'}$, $m_\alpha$, $m_\beta$, $m_\sigma$ and the quantities $q^2$ and $(p_2+p_3)^2$, which we take advantage of. We calculate the squared magnitude $|{\cal M}_{\alpha\to\beta\sigma\sigma}|^2$, then average over the initial spin $s_1$ and sum over final-state spins $s_2$, $s_3$, $s_4$. Then we define $F_\Omega=\sum_jv_{j\beta}^*v_{j\alpha}f^j_\Omega$ and use Eq.~(\ref{emvpar}) to write
\begin{eqnarray}
|\overline{\cal M}_{\alpha\to\beta\sigma\sigma}|^2&=&\frac{1}{2}\sum_{s_1=\pm}\sum_{s_2=\pm}\sum_{s_3=\pm}\sum_{s_4=\pm}|{\cal M}_{\alpha\to\beta\sigma\sigma}|^2
\nonumber \\ 
&=&\frac{e^4}{2(q^2)^2}\Big[ g^{\alpha\beta}_1|F_Q|^2+g^{\alpha\beta}_2|F_A|^2
+g^{\alpha\beta}_3|F_M|^2
+g^{\alpha\beta}_4|F_E|^2
+g^{\alpha\beta}_5{\rm Re}\{ F_QF_M^* \}
+g^{\alpha\beta}_6{\rm Im}\{ F_AF_E^* \} \Big].
\label{mnsqdamp}
\end{eqnarray}
Here, the coefficients $g^{\alpha\beta}_k$, whose explicit expressions are provided in Appendix~\ref{gApp}, depend only on the external-lepton masses $m_\alpha$, $m_\beta$, $m_\sigma$, and on the scalar products $q^2$ and $(p_2+p_3)^2$.

\subsection{The decay rate of $l_\alpha\to l_\beta \,l_\sigma \,l_\sigma$}
Now we aim at the computation of the decay rate $\Gamma_{\alpha\to\beta\sigma\sigma}=\Gamma(l_\alpha\to l_\beta \,l_\sigma \,l_\sigma)$, for which we emphasize that the dependence of $|\overline{\cal M}_{\alpha\to\beta\sigma\sigma}|^2$ on external momenta occurs exclusively through the scalar products $q^2$ and $(p_2+p_3)^2$. With that in mind, and recalling our assumption that for large $\Lambda$  the mass $m_{W'}$ is practically proportional to this scale, it is convenient to perform the changes of variables $x=q^2/m_{W'}^2$ and $y=(p_2+p_3)^2/m_{W'}^2$. We also define the squared ratios $x_\alpha=m_\alpha^2/m_{W'}^2$, $x_\beta=m_\beta^2/m_{W'}^2$, and $x_\sigma=m_\sigma^2/m_{W'}^2$. Then, after some integrations, we express the decay rate as
\begin{equation}
\Gamma_{\alpha\to\beta\sigma\sigma}=\frac{2m_{W'}}{(8\pi\sqrt{x_\alpha})^3}\int^{x_{\rm max}}_{x_{\rm min}}dx\int_{y_{\rm min}}^{y_{\rm max}}dy\,|\overline{\cal M}_{\alpha\to\beta\sigma\sigma}|^2,
\label{decayrate}
\end{equation}
with the $x$-integration limits $x_{\rm \min}=4x_\sigma$ and $x_{\rm max}=(\sqrt{x_\alpha}-\sqrt{x_\beta})^2$. On the other hand, we find that the integration limits for the $y$-integral are
\begin{eqnarray}
y_{\rm min}&=&\frac{(x_\alpha+x_\beta+2x_\sigma-x)x-\sqrt{x(x-4x_\sigma)((\sqrt{x_\alpha}+\sqrt{x_\beta})^2-x)((\sqrt{x_\alpha}-\sqrt{x_\beta})^2-x)}}{2x},
\\ \nonumber \\
y_{\rm max}&=&\frac{(x_\alpha+x_\beta+2x_\sigma-x)x+\sqrt{x(x-4x_\sigma)((\sqrt{x_\alpha}+\sqrt{x_\beta})^2-x)((\sqrt{x_\alpha}-\sqrt{x_\beta})^2-x)}}{2x}.
\end{eqnarray}
Since the factors $|F_\Omega|^2$ are independent of $y$, the integration over this variable only affects the coefficients $g_k^{\alpha\beta}$, so the $y$-integral turns out to be simple to solve. \\

\subsection{Decoupling of new physics}
For the next step, we consider a specific spectrum of heavy-neutrino masses $m_j$. The simplest choice would be to assume that the set of neutrino masses is quasi-degenerate, that is $m_1\approx m_2\approx m_3$, but doing so introduces a strong suppression of the contribution. A more interesting and flexible scenario, which we choose instead, is a mass spectrum in which two neutrinos, say $N_2$ and $N_3$, have quasi-degenerate masses, that is, $m_2\approx m_3$, but the remaining neutrino mass, $m_1$, is not close to them at all: $m_1\ne m_2$ and $m_1\ne m_3$. This neutrino-mass spectrum was considered in Ref.~\cite{NSTV} to investigate one-loop contributions to electric dipole moments and anomalous magnetic moments of Standard Model leptons, and the lepton decay $\mu\to e\gamma$ as well. We consider, for practical purposes, a mass $m_N$, such that $m_N\approx m_2$ and $m_N\approx m_3$, to characterize the pair of quasi-degenerate neutrino masses.\\

We reasonably assume that the quantities $v_{j\alpha}$, which introduce effects of heavy-neutrino mixing into the charged currents of Eq.~(\ref{CCs}), are the entries of a $3\times3$ matrix that is approximately unitary, meaning that $\sum_{j=1}^3v_{j\beta}^*v_{j\alpha}=0$ if $\alpha\ne\beta$ and $\sum_{j=1}^3v_{j\alpha}^*v_{j\alpha}\approx1$. 
Charged currents with the ingredients of heavy-neutrino mixing and a gauge boson $W'$ were utilized in a general model-independent manner in Ref.~\cite{HLRZ} to explore the observability at the Large Hadron Collider of such heavy particles through physical processes allowed in the case of heavy Majorana neutrinos. In Ref.~\cite{DBP} collider phenomenology of neutrino physics has been reviewed, considering a variety of new-physics models; this work includes bounds on the mixing of heavy neutrinos. In Ref.~\cite{GoKo}, the mixing of a new sterile neutrino, with mass within the range $10\,{\rm eV}$-$1\,{\rm TeV}$, and the Standard-Model flavor neutrinos was analyzed, under the assumption that the decay products of this heavy neutrino cannot be measured by experiments. This reference combines experimental constraints on diverse physical processes to achieve bounds on parameters of such a mixing. \\

It is worth commenting that, in general, any factor $f^j_\Omega$, in Eq.~(\ref{emvpar}), can be expressed as $f^j_\Omega=\lambda_\Omega+\cdots$, where $\lambda_\Omega$ is a term that is independent of the neutrino mass $m_j$ and the ellipsis represents those terms that, on the other hand, are $m_j$ dependent. According to the unitarity property of $v_{j\alpha}$, notice that $F_\Omega=\sum_{j=1}^3v_{j\beta}^*v_{j\alpha}(\lambda_\Omega+\cdots)=\sum_{j=1}^3v_{j\beta}^*v_{j\alpha}(\cdots)$. In other words, the adequate calculation of the contributions requires to take into account a proper implementation of the unitarity of $v_{j\alpha}$, which involves the elimination of those terms that are independent of the masses $m_j$ in any form factor $F_\Omega$. Such a correct usage of unitarity is a piece of a delicate balance that ensures the decoupling of the new-physics contribution\footnote{Choices of the gauge exist in which this property of unitarity of $v_{j\alpha}$ may even be an essential part of an exact and consistent cancellation of ultraviolet divergences, taking place through a Glashow-Iliopoulos-Maiani mechanism~\cite{GIM}.}. We have verified that performing the afore-described elimination of $m_j$-independent terms renders each heavy-neutrino contribution, separately, non-decoupling. Nevertheless, as we show in a moment, the sum of all the neutrino contributions, with this removal of $m_j$-independent terms, does decouple. We found it practical to express the form factors $F_\Omega$, in terms of the Passarino-Veltman scalar functions of Eqs.~(\ref{defB1}) to (\ref{defC0}) and $B^j_0=1$, as
\begin{eqnarray}
F_\Omega&=&\frac{v_{1\beta}^*v_{1\alpha}}{m_{W'}^{n_{\Omega}}}
\Big[
\sum_{k=0}^4\big( h^1_{\Omega k}B^1_k-h^N_{\Omega k}B^N_k \big)
+m_{W'}^2(h^1_\Omega C^1_0-h^N_\Omega C^N_0)
\Big],
\label{mjindelim}
\end{eqnarray}
with $n_Q=0$, $n_A=2$, $n_M=1$, and $n_E=1$. The coefficients $h^j_{\Omega k}$, in Eq.~(\ref{mjindelim}), are straightforwardly obtained from the form-factor contributions $f^j_{\Omega,0}$, $f^j_{\Omega,nm}$, and $f^j_{\Omega,5}$ that constitute Eq.~(\ref{finfs}). Written in this form, the elimination of $m_j$-independent terms from the factors $F_\Omega$ is automatically carried out. For the subsequent steps of this calculation, we consider that the charged-lepton masses $m_\alpha$, $m_\beta$, $m_\sigma$ are tiny with respect to the $W'$ mass, $m_{W'}$. We then use approximate solutions of the scalar functions $B_0$ and $C_0$ that are featured in our expressions. \\

Besides ultraviolet finiteness and electromagnetic gauge invariance, a further check of consistency of our results is {\it decoupling}~\cite{AppCa}. It is worth mentioning that the authors of Refs.~\cite{IlPi,GrLa} pointed out that, in the presence of heavy neutrinos, violations of the decoupling theorem may arise in flavor-lepton-violating decays of charged leptons into three charged leptons. This is not the case of our calculation. We start from Eq.~(\ref{mjindelim}), and then, using the condition $m_{W'}\approx\kappa_jm_j$, we find that any factor $F_\Omega$ can be written in the form
\begin{equation}
F_\Omega=v_{1\beta}^*v_{1\alpha}\Big[\, \frac{1}{m_{W'}^2}\eta_{\Omega1}+\frac{1}{m_{W'}^4}\eta_{\Omega2}+\frac{1}{m_{W'}^6}\eta_{\Omega3} \Big].
\label{decFs}
\end{equation}
The expressions of the coefficients $\eta_{\Omega k}$ are enormous and quite intricate, so we do not exhibit them explicitly. These coefficients only depend on the external-lepton masses and on the quantities $\kappa_1$, $\kappa_N$ and $q^2$, which are constant with respect to the energy scale $\Lambda$, for very large $\Lambda$. Since, on the other hand, $m_{W'}$ is proportional to $\Lambda$, we note, from Eq.~(\ref{decFs}), that the amplitude ${\cal M}_{\alpha\to\beta\sigma\sigma}$, given by Eq.~(\ref{invamp}), decouples as $\Lambda\to\infty$: $\lim_{\Lambda\to\infty}F_\Omega=0$.


\section{The decays $\mu\to 3e$ and $\tau\to l_\beta l_\sigma l_\sigma$}
\label{DandR}
In this section, we calculate and analyze the leading contributions from the heavy neutrinos $N_j$ and the $W'$ charged boson, in the charged currents of Eq.~(\ref{CCs}), to decays of the Standard-Model charged leptons $\mu^-$ and $\tau^-$ into three charged leptons. This comprehends the decay of the muon into $e^-e^+e^-$ and the decays of the tau lepton into $e^-\mu^+\mu^-$, $\mu^-e^+e^-$, $e^-e^+e^-$, and $\mu^-\mu^+\mu^-$. Any high-energy field formulation that intends to extend the Standard Model introduces new dynamic variables and symmetries, which naturally come along with all brand new observables to be looked for in experiments. While such new-physics effects are interesting, the physical descriptions beyond the Standard Model also affect, in general, those observables that characterize the low-energy theory, leaving traces in the form of deviations from the Standard-Model predictions. The pursuit, quantification and analysis of such effects would provide, in case of positive measurement, hints pointing towards new physics outside the reach of the Standard Model. This is the case of the lepton trilinear decays that we discuss in the present investigation, which, even involving initial and final states of low-energy dynamic variables, cannot happen in the Standard Model, where lepton flavor is preserved. \\

To compute the corresponding branching ratios, by means of Eq.~(\ref{decayrate}), we first observe, from Eqs.~(\ref{mjindelim}) and (\ref{decFs}), that the neutrino-mixing dependence of the form factors $F_\Omega$ is determined by which leptons $l_\alpha$ and $l_\beta$ we consider. A difference between Majorana and Dirac neutrinos lies in the parameters featured in the mixing. Concretely, the number of $CP$-violating phases is different~\cite{GK}: while Dirac neutrinos are characterized by only one phase, known as the {\it Dirac phase}, the case of Majorana neutrinos includes this Dirac phase and two extra $CP$-violating phases, commonly refered to as {\it Majorana phases}. Differently from what happens with the Majorana phases, neutrino oscillations are sensitive to the Dirac phase and data are available~\cite{T2K}. Consider, for the mixing of the heavy neutrinos, the usual parametrization of $3\times3$ unitary matrices~\cite{GK}, in terms of three mixing angles $\theta_{12}$, $\theta_{23}$, $\theta_{13}$, the Dirac phase $\delta$, and two Majorana phases, to write down the following mixing factors:
\begin{itemize}
\item $\mu^-\to e^-e^+e^-$:
\begin{equation}
v_{1e}^*v_{1\mu}=c_{12}c_{13}\big( -s_{12}c_{23}-c_{12}s_{23}s_{13}\,e^{-i\delta} \big),
\label{angularfactor1}
\end{equation}
\item $\tau^-\to e^-\mu^+\mu^-$ and $\tau^-\to e^-e^+e^-$:
\begin{equation}
v_{1e}^*v_{1\tau}=c_{12}c_{13}\big( s_{12}s_{23}-c_{12}c_{23}s_{13}\,e^{-i\delta} \big),
\label{angularfactor2}
\end{equation}
\item $\tau^-\to\mu^- e^+e^-$ and $\tau^-\to\mu^-\mu^+\mu^-$:
\begin{eqnarray}
v^*_{1\mu}v_{1\tau}=\big( -s_{12}c_{23}-c_{12}s_{23}s_{13}\,e^{i\delta} \big)
\big( s_{12}s_{23}-c_{12}c_{23}s_{13}\,e^{-i\delta} \big).
\label{angularfactor3}
\end{eqnarray}
\end{itemize}
Here, we have denoted, as usual, $\sin\theta_{kn}=s_{kn}$ and $\cos\theta_{kn}=c_{kn}$, where the $\theta_{kn}$ are the aforementioned mixing angles for heavy neutrinos. Note that the factors given in Eqs.~(\ref{angularfactor1}) to (\ref{angularfactor3}) are independent of Majorana phases. 
According to Eqs.~(\ref{mnsqdamp}), (\ref{decayrate}), and (\ref{mjindelim}), the branching ratio for $l_\alpha\to l_\beta l_\sigma l_\sigma$ (leading contributions) has the general structure ${\rm Br}(l_\alpha\to l_\beta l_\sigma l_\sigma)=|v^*_{1\beta}v_{1\alpha}|^2(\cdots)$, where $(\cdots)$ represents some function that is independent of mixing parameters. Moreover, this illustrates that rather than the sole factors $v_{1\beta}^*v_{1\alpha}$, we need the factors $|v_{1\beta}^*v_{1\alpha}|^2$, in order to calculate the branching ratios for these processes. It turns out that $0\leqslant|v^*_{1\beta}v_{1\alpha}|^2 \leqslant 1/4$ for any $\alpha$ and $\beta$, that is, the neutrino-mixing dependence of the branching ratios has an upper bound. \\

The branching ratios for the processes that we consider have been reported to have the following upper bounds~\cite{PDG,SINDRUM,BABARLFVtau,BelleLFVtau}:
\begin{eqnarray}
{\rm Br}(\mu^-\to e^-e^+e^-)<1.0\times10^{-12},
\label{brmtoeee}
\\ \nonumber \\
{\rm Br}(\tau^-\to e^-\mu^+\mu^-)<2.7\times10^{-8},
\label{brtautoemm}
\\ \nonumber \\
{\rm Br}(\tau^-\to \mu^-e^+e^-)<1.8\times10^{-8},
\label{brtautomee}
\\ \nonumber \\
{\rm Br}(\tau^-\to e^-e^+e^-)<2.7\times10^{-8},
\label{brtautoeee}
\\ \nonumber \\
{\rm Br}(\tau^-\to \mu^-\mu^+\mu^-)<2.1\times10^{-8},
\label{brtautommm}
\end{eqnarray}
all of them at the $90\%$ C.L. The idea is to compare these upper bounds with our branching ratios, and find sets of parameters $m_{W'},\kappa_1,\kappa_N$ such that Eqs.~(\ref{brmtoeee}) to (\ref{brtautommm}) are fulfilled.\\

Neutrino mixing, which is not a feature of the Standard Model, enables processes that violate lepton flavor, such as the muon decay $\mu\to e\gamma$. This decay has been bounded by the MEG Collaboration, which reported, in Ref.~\cite{MEGcoll}, the upper limit ${\rm Br}(\mu\to e\gamma)<5.7\times10^{-13}$. The contributions to $\mu\to e\gamma$ from charged currents involving heavy neutrinos, heavy charged gauge bosons and Standard-Model charged leptons, in a model-independent context, were recently calculated, analyzed and discussed in Ref.~\cite{NSTV}. The authors of Ref.~\cite{AADDQV} performed a calculation of ${\rm Br}(\mu\to e\gamma)$ from the same type of charged currents, in a non-supersymmetric model aimed at the stabilization of dark matter. The authors of Ref.~\cite{Boltonetal} performed an experimental search based on an apparatus that they refer to as the ``Crystal Box'' detector, aiming at the measurement of the lepton-flavor-violating decays $\mu\to e\gamma$, $\mu\to e\gamma\gamma$, and $\mu\to 3e$. Having not found any evidence of such processes, they established upper bounds, particularly ${\rm Br}(\mu\to 3e)<3.5\times10^{-11}$ at 90\% C.L. An improved bound was provided by the SINDRUM Collaboration, which, using a large-solid-angle magnetic spectrometer, achieved ${\rm Br}(\mu\to 3e)<1.0\times10^{-12}$ in Ref.~\cite{SINDRUM}. While these experimental investigations of $\mu\to 3e$ were carried out long ago, another experiment, called ``Mu3e", has been proposed with the objective of improving this lower bound to $10^{-16}$~\cite{Mu3ecoll}. Phenomenological investigations of the decay $\mu\to 3e$ have been performed within models of several Higgs doublets, leading to branching ratios as large as $\sim10^{-13}$~\cite{GrLa,AGL}. This process has been explored in the context of left-right symmetric models as well\footnote{Plentiful and useful information about lepton-flavor violation in left-right symmetric models, including these muon decays, can be found in Ref.~\cite{BaRo}.}, finding that contributions around $\sim10^{-16}-10^{-12}$ can be achieved~\cite{BoKu}. The so-called {\it scotogenic model}~\cite{ErnestMa}, aiming at neutrino mass and dark matter, has also been considered, in Ref.~\cite{ToVi}, to study this flavor-lepton-violating process, with values of ${\rm Br}(\mu\to 3e)$ ranging from $\sim10^{-18}$ to $\sim10^{-4}$ reported for a variety of scenarios. \\

\subsection{Restrictions on $m_{W'}$ from $\mu\to3e$}
\begin{figure}[!ht]
\center
\includegraphics[width=5.9cm]{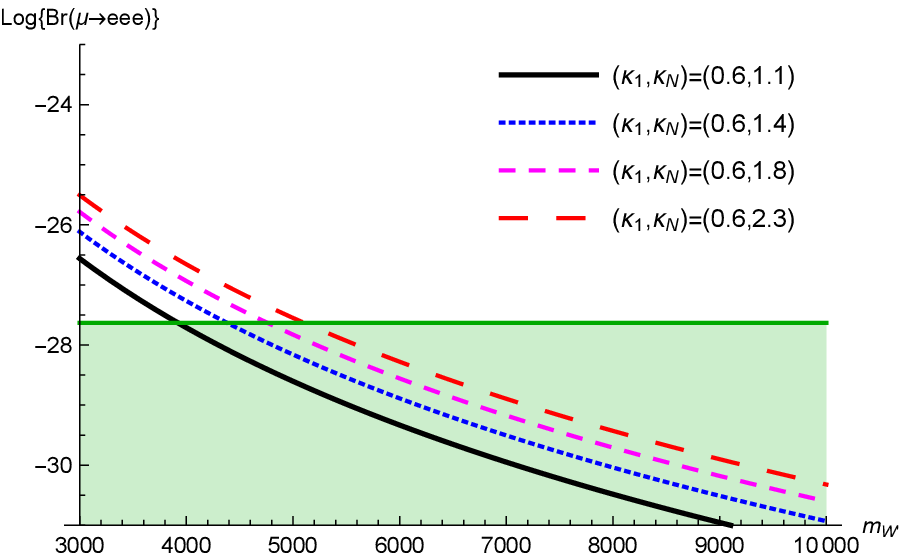}
\includegraphics[width=5.9cm]{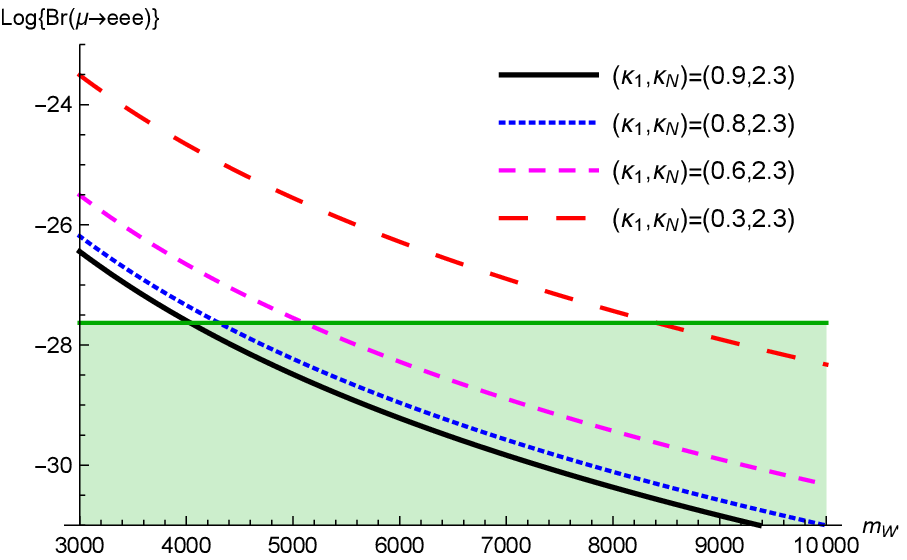}
\includegraphics[width=5.9cm]{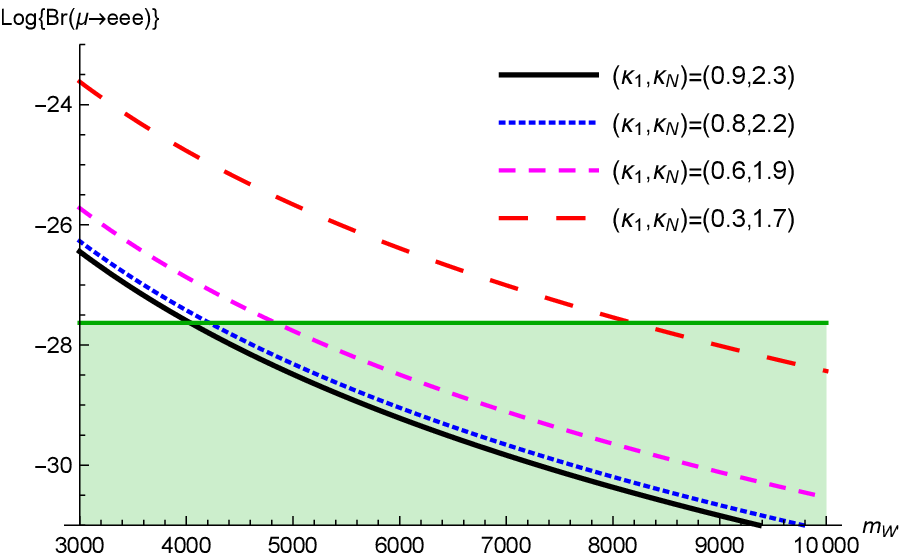}
\\ Graph (1) \hspace{4.5cm} Graph (2) \hspace{4.5cm} Graph (3)
\caption{\label{ksinmtoeee} Contributions to ${\rm Br}(\mu\to3e)$ as a function of $m_{W'}$ (in GeV units), for different $(\kappa_1,\kappa_N)$, with $\kappa_1<1<\kappa_N$: (1) constant $\kappa_1$ and increasing $\kappa_N$; (2) constant $\kappa_N$ and decreasing $\kappa_1$; decreasing $\kappa_1$ and $\kappa_N$, but constant $\kappa_1-\kappa_N$. The graphs have been plotted in logarithmic scale.}
\end{figure}

Concerning our analysis, we have observed that, by far, the decay $\mu\to 3e$ is the process which imposes the most stringent restrictions among the lepton-flavor-violating trilinear decays under consideration. In Fig.~\ref{ksinmtoeee}, we show three sets of plots of ${\rm Br}(\mu\to3e)$, in logarithmic scale, as a function of $m_{W'}$. Each curve corresponds to a different choice of pairs of parameters $(\kappa_1,\kappa_N)$. 
As we mentioned before, the dependence of the branching ratios on the mixing of heavy neutrinos is contained within factors $|v^*_{j\beta}v_{j\alpha}|^2$. To plot the graphs of Fig~\ref{ksinmtoeee},
we have used the upper bound of such a neutrino-mixing dependence ($0\leqslant|v^*_{1e}v_{1\mu}|^2 \leqslant 1/4$), which means that each graph is, in this sense, an upper bound of a branching ratio, so the region below it comprises all the possible values that correspond to the different sets of mixing parameters.
In all cases we have considered masses $m_{W'}$ ranging between $3\,{\rm TeV}$ and $10\,{\rm TeV}$. In each of these graphs, we have included a solid horizontal line that corresponds to ${\rm Br}(\mu\to3e)=1.0\times10^{-12}$, which stands for the upper bound given in Eq.~(\ref{brmtoeee}). So, the shadowed region, below such a line, comprises all the values that are allowed by the aforementioned constraint. For Graph (1) we have left $\kappa_1$ without any change, while $\kappa_N$ has been varied; it can be appreciated that as we take larger values of $\kappa_N$, meaning that $m_N$ decreases and the difference $|m_1-m_N|$ increases, the contribution gets larger as well. We have done something different to set the situation shown in Graph (2), where $\kappa_N$ is kept constant, but different values of $\kappa_1$ are taken. We see that the smaller the $\kappa_1$, which yields larger masses $m_1$ and larger differences $|m_1-m_N|$, the larger the contribution. Finally, in Graph (3), we have plotted the branching ratio for different values of $(\kappa_1,\kappa_N)$, but keeping the difference among the kappas constant. From this last graph, it is clear that the contribution grows as both $m_1$ and $m_N$ increase, even though $|m_1-m_N|$ remains constant. Besides these patterns, the graphs in Fig~(\ref{ksinmtoeee}) show us that for each pair of values $(\kappa_1,\kappa_N)$ there is a minimum value of the $W'$ mass that keeps the contribution under the SINDRUM upper bound given in Eq.~(\ref{brmtoeee}). Let us point out that, though with different numbers, all this discussion on the behavior of the branching ratio is similar for the tau decays $\tau\to l_\beta\, l_\sigma\, l_\sigma$. \\

\begin{figure}[!ht]
\includegraphics[width=8.6cm]{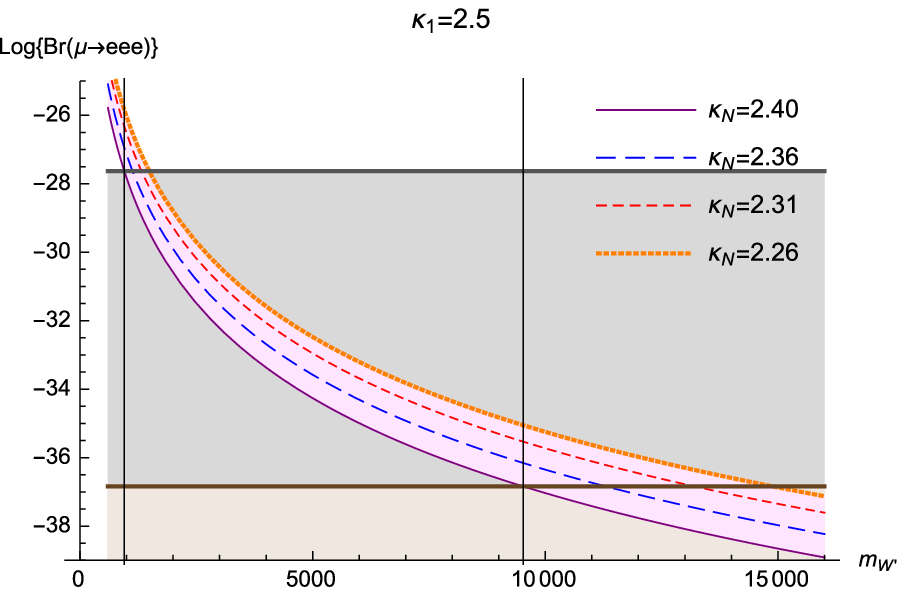}
\includegraphics[width=8.6cm]{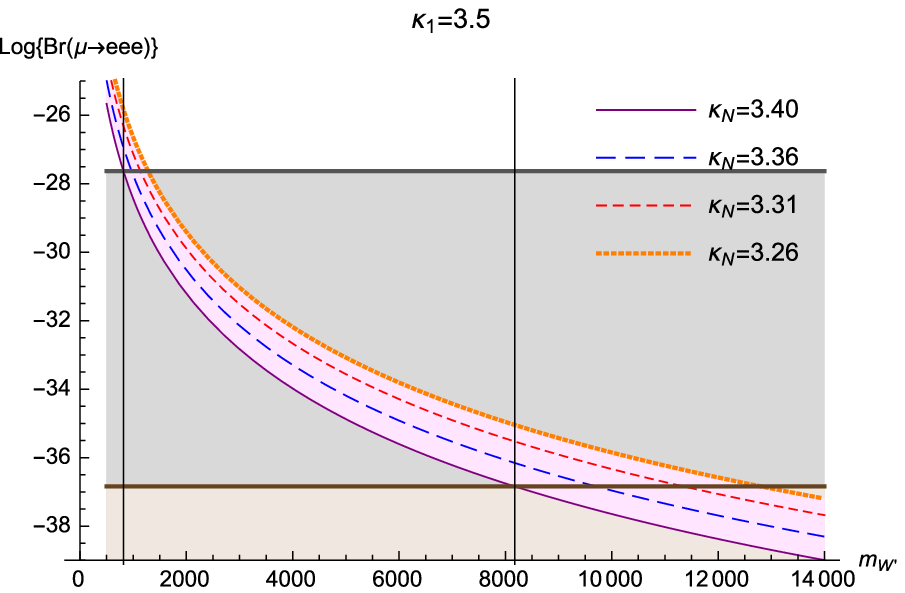}
\\
Graph (a)\hspace{7cm} Graph (b)
\\
\vspace{0.5cm}
\includegraphics[width=8.6cm]{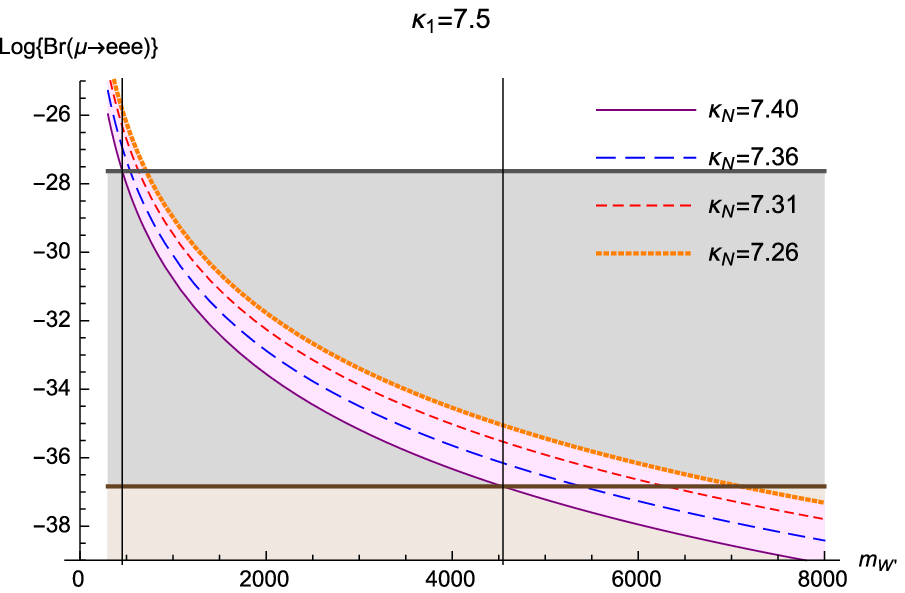}
\includegraphics[width=8.6cm]{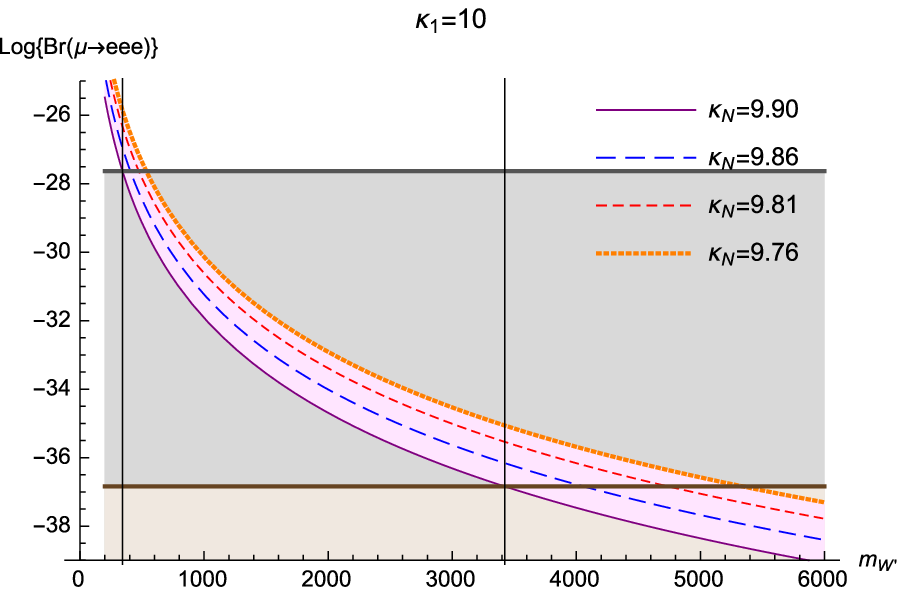}
\\
Graph (c)\hspace{7cm} Graph (d)
\caption{\label{graphsforbounds} Improvement of lower bounds on the $W'$ mass through suitable choices of the kappa parameters $\kappa_1$ and $\kappa_N$: increased kappas, corresponding to smaller heavy-neutrino masses; and kappas lying closer to each other. The lowest lower bounds on $m_{W'}$, which are indicated in each graph by a vertical solid line, are given in Table~\ref{mwboundsandnumasses}. In all cases, mass units are GeVs.}
\end{figure}

For the sets of parameters $(\kappa_1,\kappa_N)$ that we considered for Fig.~\ref{ksinmtoeee}, the lowest lower bound on $m_{W'}$ is large: $m_{W'}\gtrsim4\,{\rm TeV}$. 
Keep in mind that here and in what follows ``lower bound'' refers to the smallest allowed value of the $W'$ mass for the largest neutrino-mixing dependence, which must be emphasized because different sets of neutrino-mixing parameters would decrease this value further.
From the discussion of the previous paragraph, it seems that we can reduce the minimal $m_{W'}$ by taking increasing values of $\kappa_1$ and $\kappa_N$ that are, in addition, close to each other. This is what we have done to carry out the graphs of Fig.~\ref{graphsforbounds},
which exhibit, in logarithmic scale, sets of plots of ${\rm Br}(\mu\to3e)$ as a function of $m_{W'}$, for different choices of kappa parameters $(\kappa_1,\kappa_N)$. To perform a clear discussion, we have set, in each graph, a fixed $\kappa_1$, but we have explored different values of $\kappa_N$. 
We have taken the largest neutrino-mixing factor to get these plots. All the graphs include two horizontal lines: the upper line corresponds to the SINDRUM~\cite{SINDRUM} upper bound on ${\rm Br}(\mu\to3e)$, while the lower line represents the expected improvement of this bound by the Mu3e experiment~\cite{Mu3ecoll}. The regions below each of these lines give the corresponding allowed values for ${\rm Br}(\mu\to3e)$.
The situation of Graph (a), for which $\kappa_1=2.5$, is able to provide a lower bound for the $W'$ mass as small as $m_{W'}\gtrsim952.84\,{\rm GeV}$, which we have indicated in the corresponding image by means of the left vertical solid line. This lower bound for $m_{W'}$ improves that of Fig.~\ref{ksinmtoeee}. We can further reduce this lower bound. Graph (b) shows plots of the branching ratio with larger values of $\kappa_1$ and $\kappa_N$. Since this decreases the heavy-neutrino masses $m_1$ and $m_N$, the contributions are rendered smaller, and thus the lowest lower bound on the $W'$ mass turns out to be $m_{W'}\gtrsim817.71\,{\rm GeV}$. For Graph (c) we have the same pattern, yielding $m_{W'}\gtrsim454.54\,{\rm GeV}$, but the lowest lower bound among all the cases that we investigated comes from Graph (d), since, in this case, the bound is as small as $m_{W'}\gtrsim342.36\,{\rm GeV}$. 
In case of no detection of $\mu\to3e$, the improved sensibility of Mu3e would increase these lower bounds on $m_{W'}$ by one order of magnitude, which has been indicated in the graphs of Fig.~\ref{graphsforbounds} by means of the right vertical lines.
In Table~\ref{mwboundsandnumasses},
\begin{table}[!ht]
\center
\begin{tabular}{|c|c|c|c|c|c|c|c|c|}
\hline
 & Graph (a) & Graph (a) & Graph (b) & Graph (b) & Graph (c) & Graph (c) & Graph (d) & Graph (d) 
\\ 
& SINDRUM & Mu3e & SINDRUM & Mu3e & SINDRUM & Mu3e & SINDRUM & Mu3e
\\ \hline
$m_{W'}^{\rm min}$ & 952.84 & 9528.35 & 817.71 & 8177.10 & 454.54 & 4545.38 & 342.36 & 3423.55
\\ 
$m_{1}^{\rm min}$ & 381.14 & 3811.34 & 233.63 & 2336.31 & 60.61 & 606.05 & 34.24 & 342.36
\\ 
$m_{N}^{\rm min}$ & 397.02 & 3970.15 & 240.50 & 2405.03 & 61.42 & 614.24 & 34.58 & 347.22
\\ 
$m_{W'}^{\rm max}$ & 1489.85 & 14898.5 & 1281.32 & 12813.20 & 710.44 & 7104.35 & 534.25 & 5342.48
\\ 
$m_{1}^{\rm max}$ & 595.94 & 5959.40 & 366.09 & 3660.91 & 94.73 & 947.25 & 53.42 & 534.24
\\ 
$m_{N}^{\rm max}$ & 659.23 & 6592.26 &  393.04 & 3930.43 & 97.86 & 978.56 & 54.74 & 547.39
\\ \hline
\end{tabular}
\caption{\label{mwboundsandnumasses} Values of $m_{W'},m_1,m_N$, from graphs of Fig.~\ref{graphsforbounds}, corresponding to the bounds from SINDRUM~\cite{SINDRUM} and the expected improvement to be introduced by Mu3e~\cite{Mu3ecoll}. The first row provides lowest lower bounds on $m_{W'}$; the second and third rows give, from the $\kappa_1$ and $\kappa_N$ associated to the column value of $m_{W'}$, the corresponding heavy neutrino masses. We have also included the maximum lower bounds on $m_{W'}$ for the sets of kappa parameters considered in each graph. All the masses are expressed in GeVs.}
\end{table}
we give, for each graph of Fig~\ref{graphsforbounds}, the lowest lower bound on $m_{W'}$ and the heavy-neutrino masses corresponding to each scenario and to the sensibilities of SINDRUM and Mu3e. \\

Let us recall and emphasize that our preceding discussion around Fig.~\ref{ksinmtoeee} indicates that, for any fixed $\kappa_1$, a smaller $\kappa_N$ would render the neutrino mass $m_N$ larger, thus increasing the difference $|m_1-m_N|$ and, consequently, enhancing the corresponding contribution. This behavior is followed by all the graphs of Fig.~\ref{graphsforbounds}, where we observe that larger differences among kappas push the lower bound on the $W'$ mass forward. For instance, decreasing $\kappa_N$ from 9.90 to 9.76 in Graph (d) increases the lower bound $m_{W'}\gtrsim342.36\,{\rm GeV}$ to $m_{W'}\gtrsim534.25\,{\rm GeV}$, with neutrino masses $m_1\approx53.96\,{\rm GeV}$ and $m_N\approx54.74\,{\rm GeV}$. \\

\begin{figure}[!ht]
\center
\includegraphics[width=8.7cm]{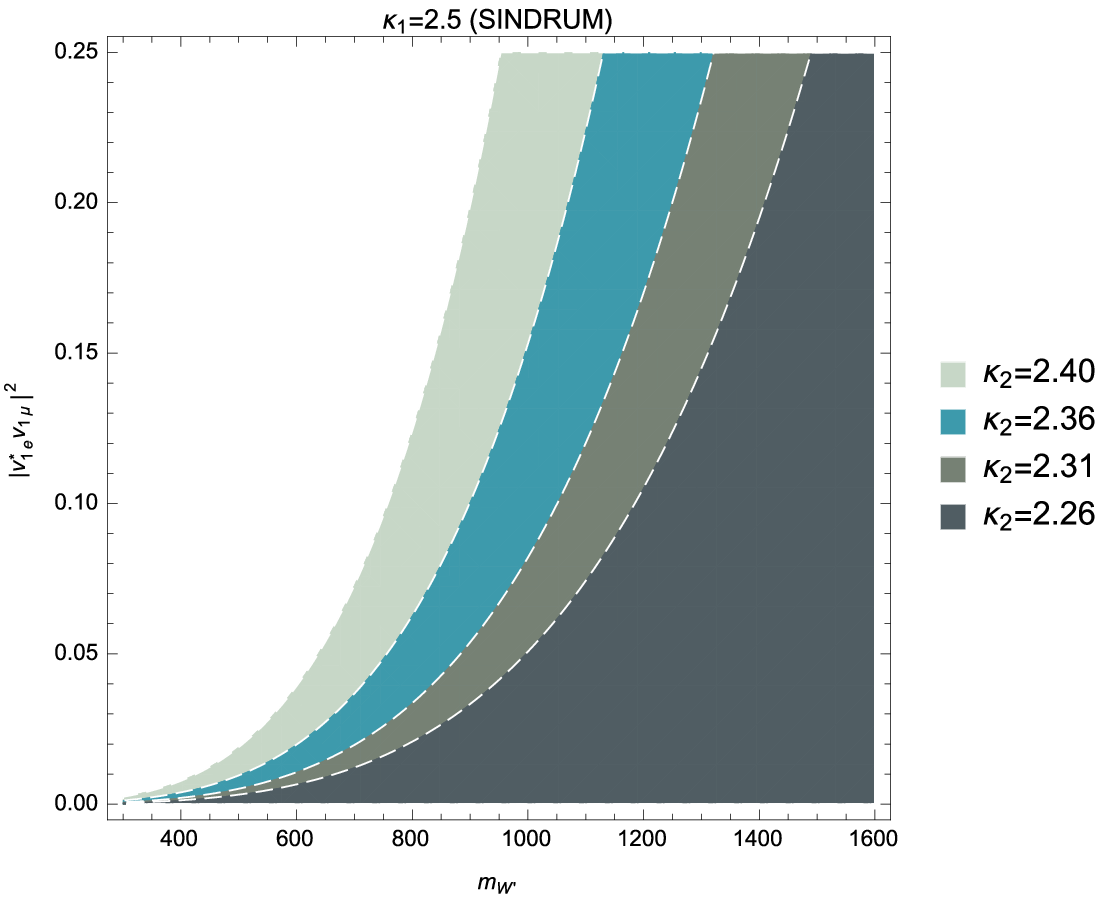}
\hspace{0.05cm}
\includegraphics[width=8.7cm]{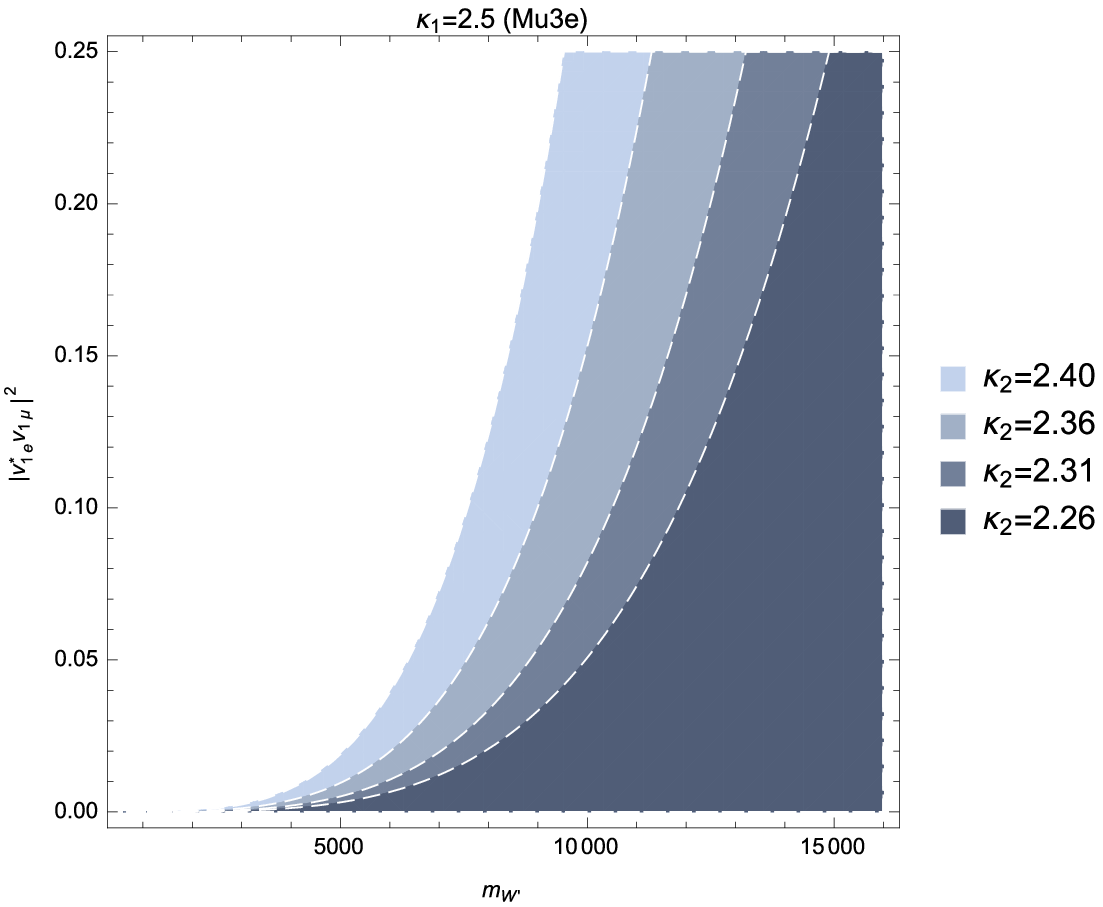}
\caption{\label{mwVSmixing1} Allowed regions for ${\rm Br}(\mu\to3e)$, in the parameter space $(m_{W'},|v^*_{1e}v_{1\mu}|^2)$, according to the SINDRUM Collaboration (left graph) and to the Mu3e expected sensitivity (right graph). The kappa parameters under consideration are $(\kappa_1=2.50,\kappa_N=2.40)$, $(\kappa_1=2.50,\kappa_N=2.36)$, $(\kappa_1=2.50,\kappa_N=2.31)$, $(\kappa_1=2.50,\kappa_N=2.26)$.}
\end{figure}

\begin{figure}[!ht]
\center
\includegraphics[width=8.7cm]{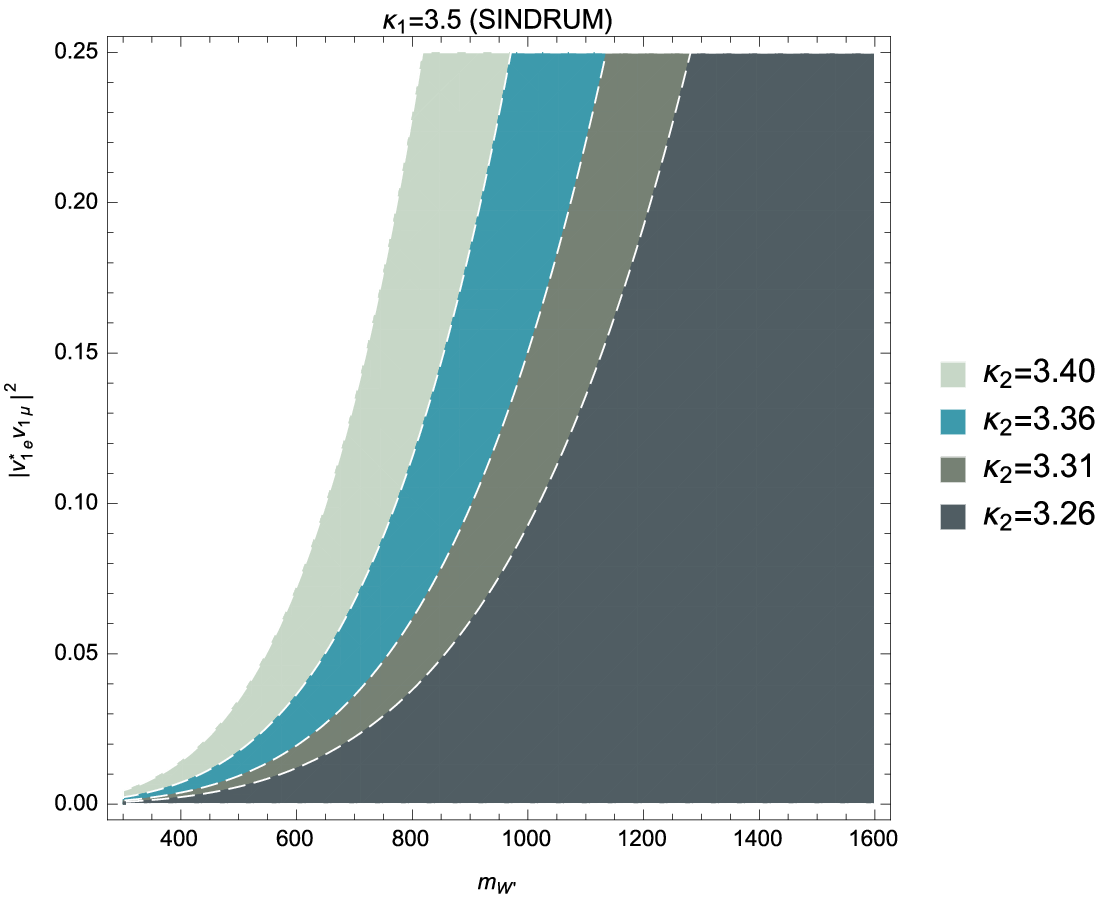}
\hspace{0.05cm}
\includegraphics[width=8.7cm]{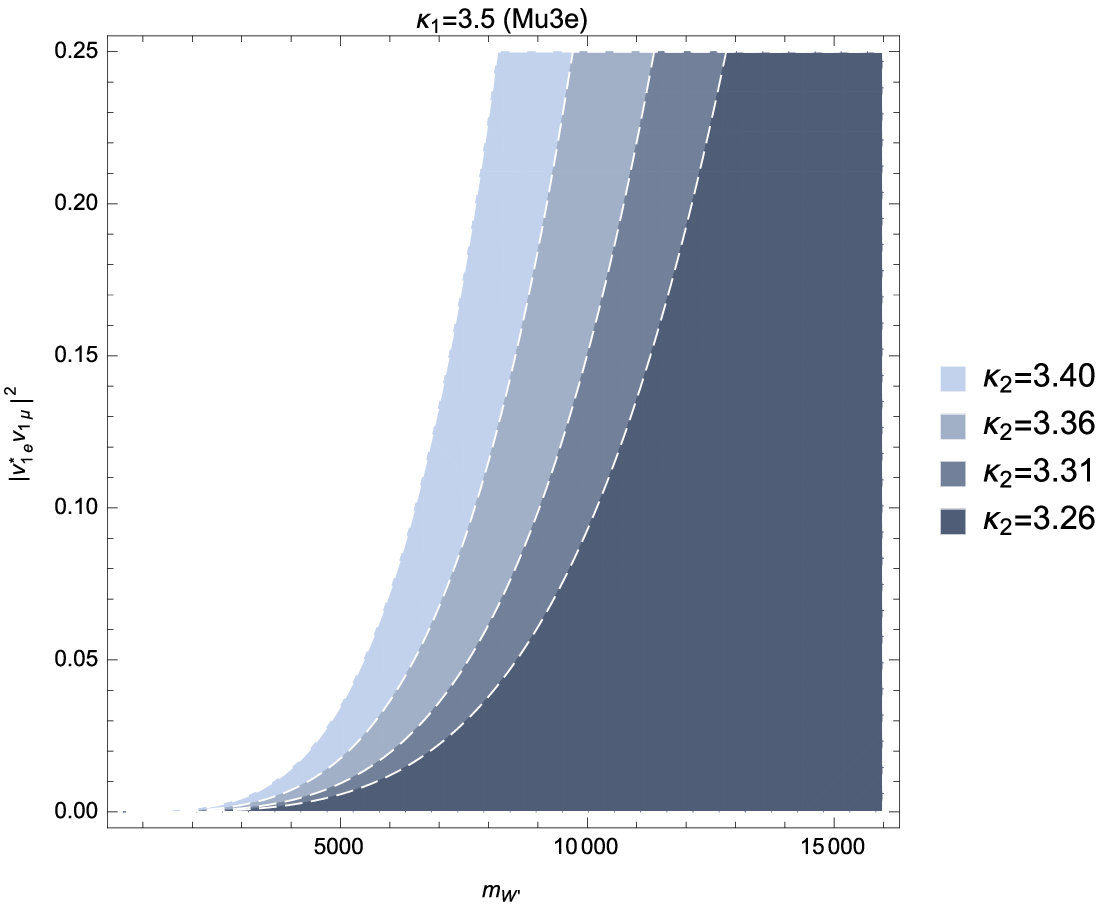}
\caption{\label{mwVSmixing2} Allowed regions for ${\rm Br}(\mu\to3e)$, in the parameter space $(m_{W'},|v^*_{1e}v_{1\mu}|^2)$, according to the SINDRUM Collaboration (left graph) and to the Mu3e expected sensitivity (right graph). The kappa parameters under consideration are $(\kappa_1=3.50,\kappa_N=3.40)$, $(\kappa_1=3.50,\kappa_N=3.36)$, $(\kappa_1=3.50,\kappa_N=3.31)$, $(\kappa_1=3.50,\kappa_N=3.26)$.}
\end{figure}

\begin{figure}[!ht]
\center
\includegraphics[width=8.7cm]{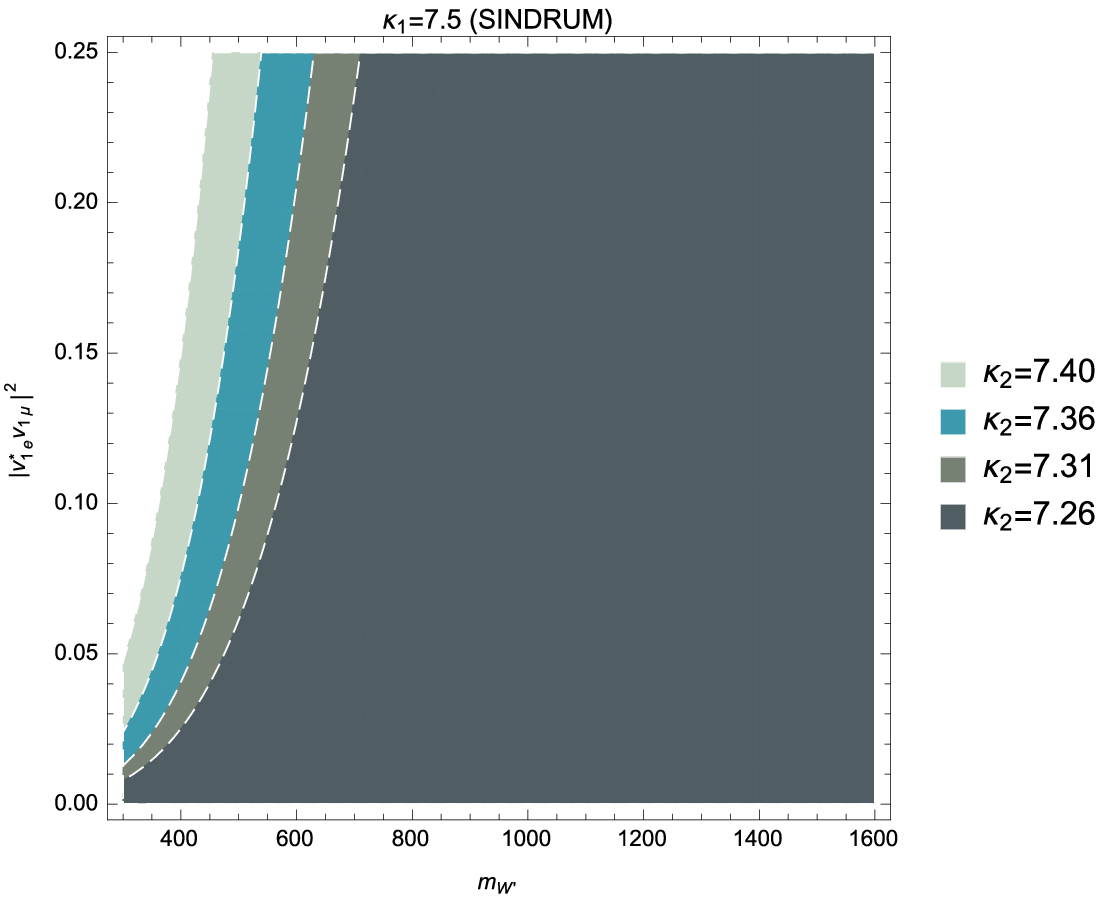}
\hspace{0.05cm}
\includegraphics[width=8.7cm]{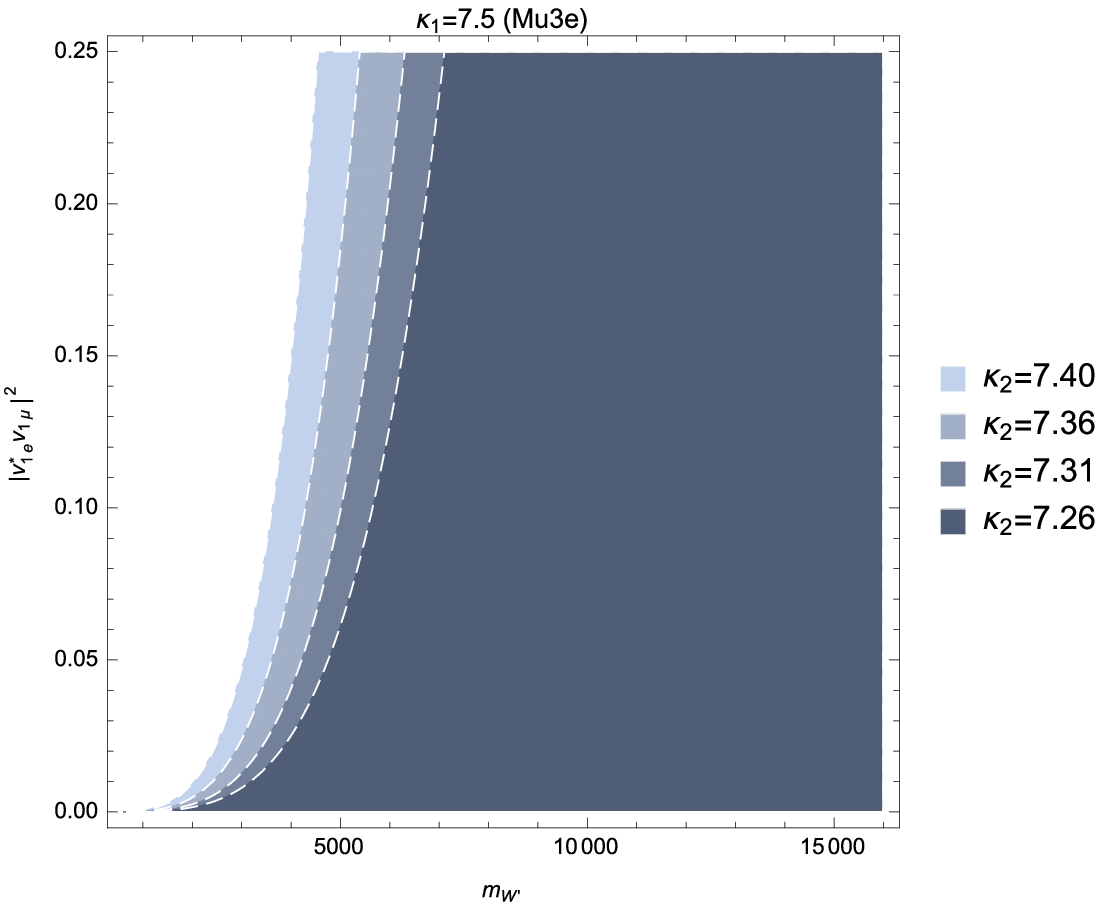}
\caption{\label{mwVSmixing3} Allowed regions for ${\rm Br}(\mu\to3e)$, in the parameter space $(m_{W'},|v^*_{1e}v_{1\mu}|^2)$, according to the SINDRUM Collaboration (left graph) and to the Mu3e expected sensitivity (right graph). The kappa parameters under consideration are $(\kappa_1=7.50,\kappa_N=7.40)$, $(\kappa_1=7.50,\kappa_N=7.36)$, $(\kappa_1=7.50,\kappa_N=7.31)$, $(\kappa_1=7.50,\kappa_N=7.26)$.}
\end{figure}

\begin{figure}[!ht]
\center
\includegraphics[width=8.7cm]{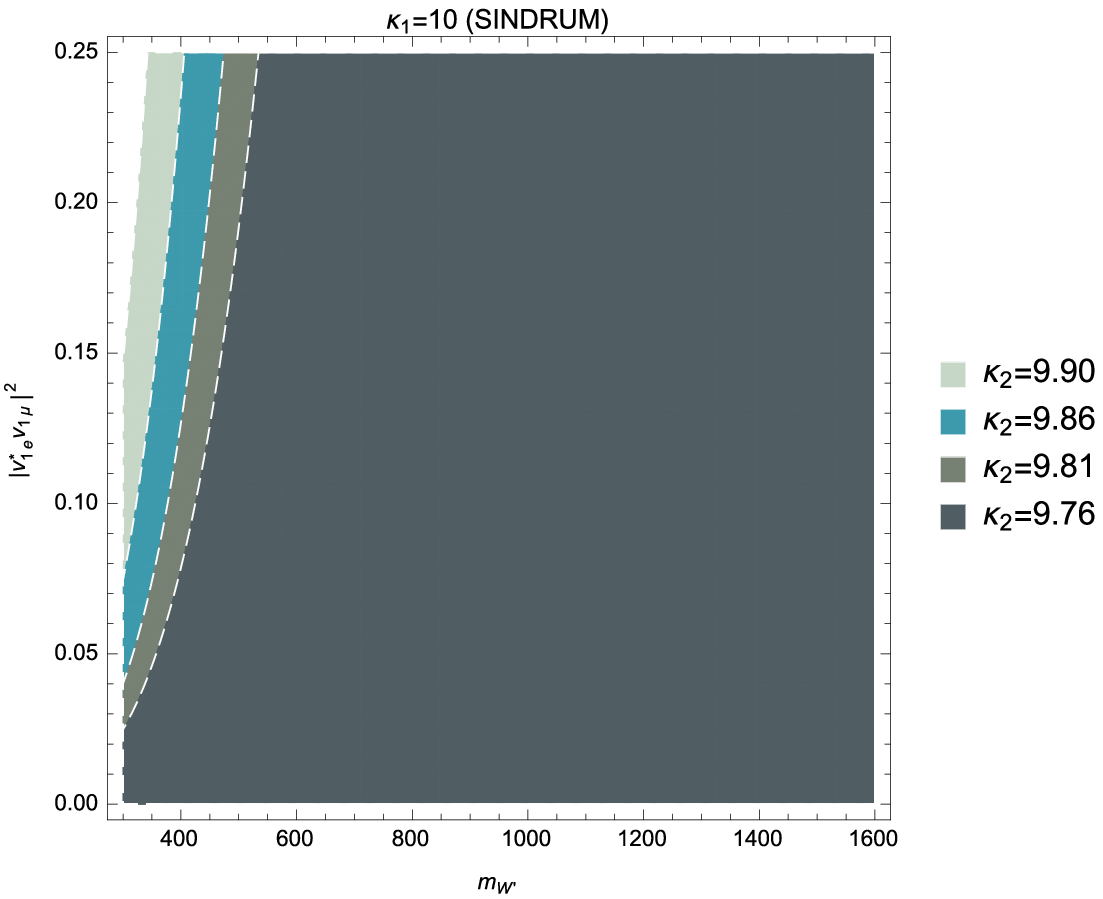}
\hspace{0.05cm}
\includegraphics[width=8.7cm]{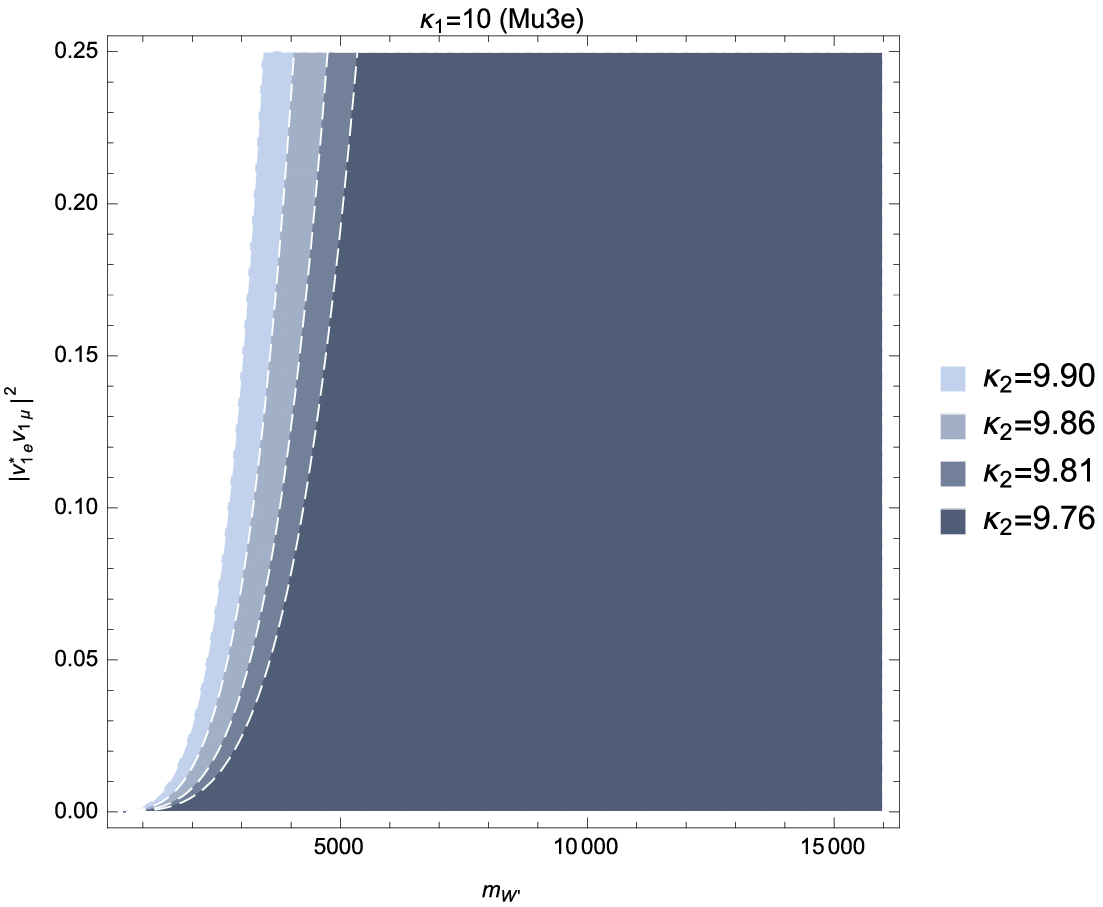}
\caption{\label{mwVSmixing4} Allowed regions for ${\rm Br}(\mu\to3e)$, in the parameter space $(m_{W'},|v^*_{1e}v_{1\mu}|^2)$, according to the SINDRUM Collaboration (left graph) and to the Mu3e expected sensitivity (right graph). The kappa parameters under consideration are $(\kappa_1=10,\kappa_N=9.90)$, $(\kappa_1=10,\kappa_N=9.86)$, $(\kappa_1=10,\kappa_N=9.81)$, $(\kappa_1=10,\kappa_N=9.76)$.}
\end{figure}

The whole neutrino-mixing dependence of ${\rm Br}(\mu\to3e)$ lies within the factor $|v^*_{1e}v_{1\mu}|^2$. With this in mind, we provide Figs.~\ref{mwVSmixing1}-\ref{mwVSmixing4}. 
Each of these figures comprises two graphs in which regions in the parameter space $(m_{W'},|v^*_{1e}v_{1\mu}|^2)$ are shown. Furthermore, each figure (with its two graphs) is associated to a fixed parameter $\kappa_1$, analogously to what we did in Fig.~\ref{graphsforbounds}. In all these figures we have taken values of the $W'$ mass ranging from $0.6\,{\rm TeV}$ to $1.6\,{\rm TeV}$ for left graphs and $0.6\,{\rm TeV}$ to $16\,{\rm TeV}$ for right graphs. The left graph of each figure shows regions in this parameter space which are allowed by the SINDRUM bound, with each region corresponding to a fixed pair $(\kappa_1,\kappa_N)$. We have used the same sets of kappa parameters as those of Fig.~\ref{graphsforbounds}. Allowed regions have been colored from lightest to darkest, while forbidden regions, on the left, are colorless. The allowed regions in a given graph are not disjoint of each other, but they spread to the right and share points. Something similar has been done with the right graphs of these figures, but with the difference that they show regions that would remain allowed by the expected sensitivity of the experiment Mu3e. According to all the graphs of Figs.~\ref{mwVSmixing1}-\ref{mwVSmixing4}, as larger values of $\kappa_1$ and $\kappa_N$, corresponding to smaller heavy-neutrino masses, are considered, the resulting allowed regions are larger, and the neutrino-mixing factor $|v^*_{1e}v_{1\mu}|^2$ is thus less restricted for small values of $m_{W'}$. The same occurs as one explores kappas $\kappa_1$ and $\kappa_N$ that are closer to each other, which translates into neutrino masses that are more alike to each other. On the contrary, smaller kappas (heavier neutrinos) and/or larger diferences among kappas yield smaller regions in which $|v^*_{1e}v_{1\mu}|^2$ would be stringently constrained in lighter-$W'$ scenarios. \\

\subsection{$\mu\to e\gamma$ VS $\mu\to3e$}
\begin{figure}[!ht]
\center
\includegraphics[width=8.6cm]{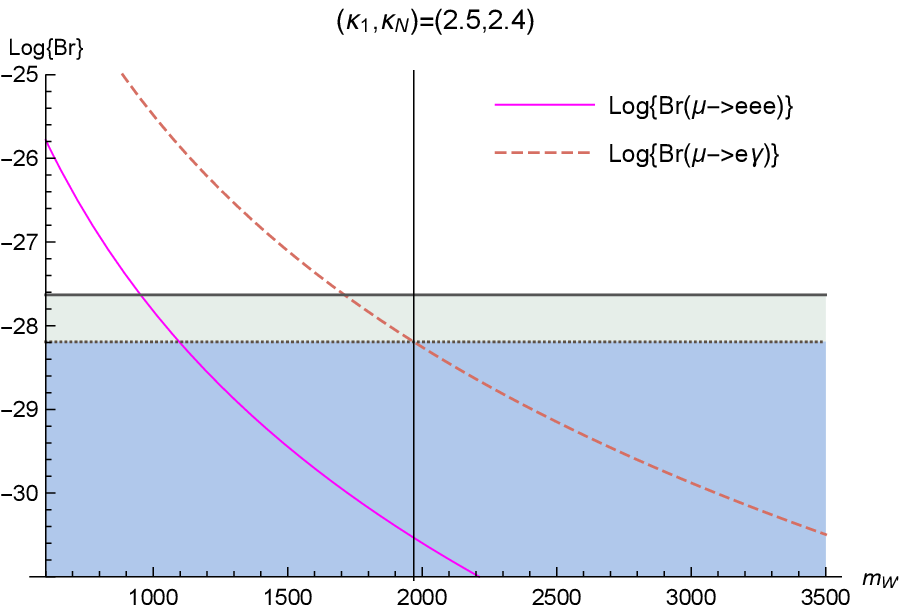}
\includegraphics[width=8.6cm]{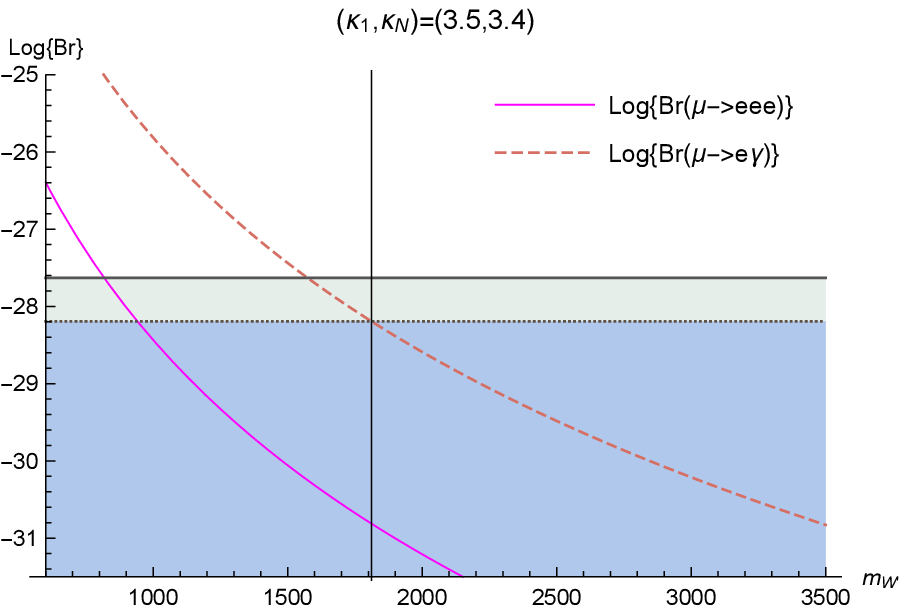}
\\
Graph (I)\hspace{7cm} Graph (II)
\\
\vspace{0.5cm}
\includegraphics[width=8.6cm]{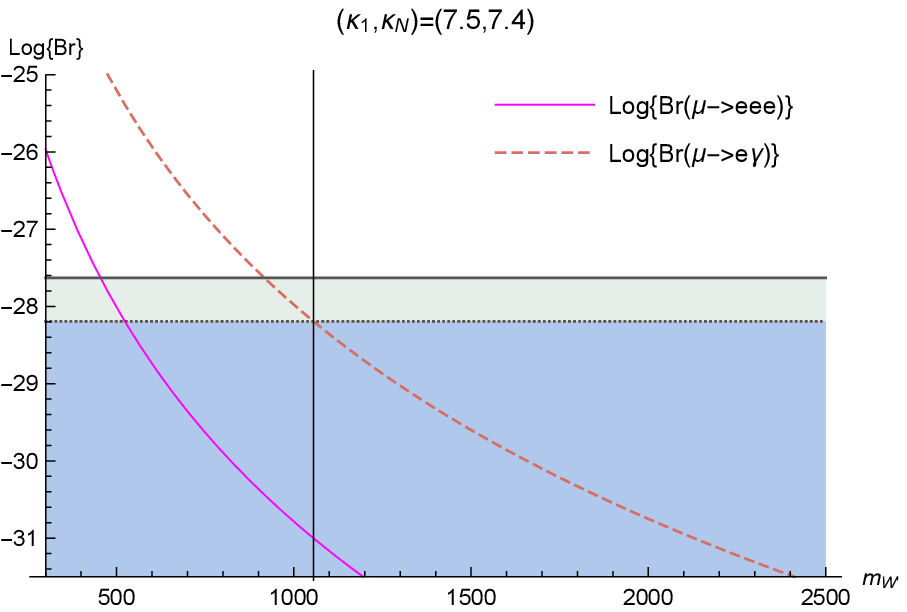}
\includegraphics[width=8.6cm]{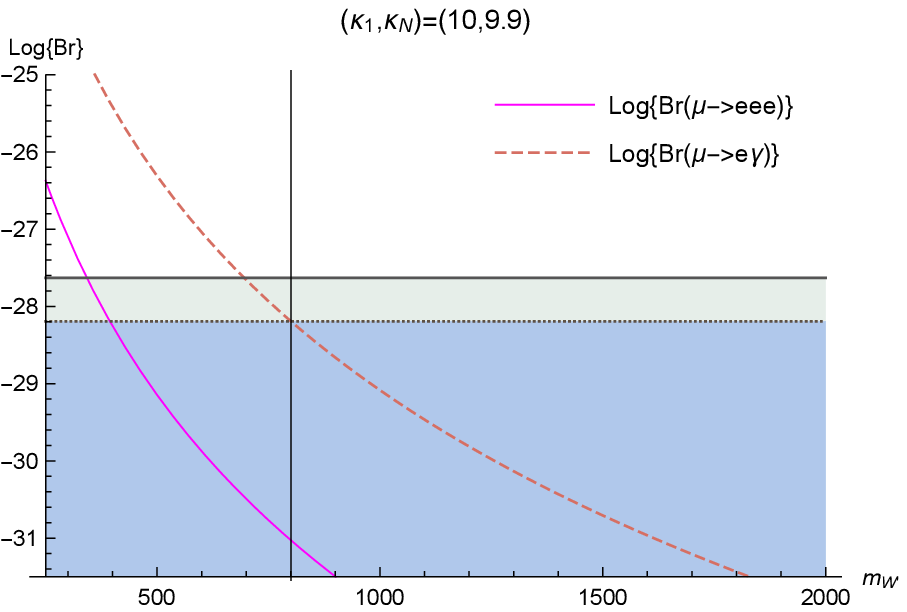}
\\
Graph (III)\hspace{7cm} Graph (IV)
\caption{\label{3evsgegraphs} Comparison of ${\rm Br}(\mu\to e\gamma)$ with ${\rm Br}(\mu\to3e)$. The branching ratio of $\mu\to e\gamma$ has been plotted as dashed curves (red), and the upper bound on this quantity corresponds to the MEG limit (dotted horizontal line). Solid curves (magenta) represent the branching ratio of $\mu\to3e$, whose allowed region is bounded from above by the SINDRUM upper limit (solid horizontal line). All the graphs have been plotted in logarithmic scale.}
\end{figure}

In the next step we compare the trilinear decay $\mu\to3e$ with the flavor-changing process $\mu\to e\gamma$. In order to do that, we use the results of Ref.~\cite{NSTV}, where the contributions from general charged currents to Standard-Model charged-lepton electromagnetic moments, both diagonal and non-diagonal, and the branching ratio of the flavor-violating decay $\mu\to e\gamma$ where calculated. The expression that we need for the decay rate $\Gamma(\mu\to e\gamma)$ is~\cite{NSTV}
\begin{eqnarray}
\Gamma(l_\mu\to l_e\gamma)&=&\frac{e^2 |v^*_{1e}v_{1\mu}|^2}{\pi}\frac{(m_\mu^2-m_e^2)^3}{m_\mu^3m_1^4m_N^4}\bigg\{ \bigg[ \frac{(m_\mu^2+m_e^2)(m_N^2\eta^{(2)}_1-m_1^2\eta^{(2)}_N)+m_\mu m_e(m_N^2\eta^{(3)}_1-m_1^2\eta^{(3)}_N)}{m_\mu+m_e} \bigg]^2 
\nonumber \\ &&
+\bigg[
\frac{(m_\mu^2+m_e^2)(m_N^2\omega^{(2)}_1-m_1^2\omega^{(2)}_N)+m_\mu m_e(m_N^2\omega^{(3)}_1-m_1^2\omega^{(3)}_N)}{m_\mu-m_e}
\bigg]^2
\bigg\},
\label{decayratemuegamma}
\end{eqnarray}
with the definitions
\begin{eqnarray}
\eta^{(2)}_{j}&=&\frac{2 \kappa _j^8-27 \kappa _j^6+32 \kappa _j^4-9 \kappa _j^2+2 \left(4\kappa _j^4+6 \kappa _j^2-1\right) \kappa _j^2 \log\kappa _j^2+2}{2(16\pi)^2 \kappa _j^2\left(\kappa _j^2-1\right){}^4},
\\ \nonumber \\
\eta^{(3)}_{j}&=&\frac{6 \kappa _j^6-29 \kappa _j^4+26 \kappa _j^2+2 \left(2 \kappa_j^2+5\right) \kappa _j^2 \log\kappa _j^2-3}{2(16\pi)^2 \kappa _j^2 \left(\kappa _j^2-1\right){}^3},
\end{eqnarray}
\begin{eqnarray}
\omega^{(2)}_{j}&=&\frac{-2 \kappa _j^8+27 \kappa _j^6-32 \kappa _j^4+9 \kappa _j^2-2 \left(4\kappa _j^4+6 \kappa _j^2-1\right) \kappa _j^2 \log\kappa _j^2-2}{2(16\pi)^2 \kappa _j^2\left(\kappa _j^2-1\right){}^4},
\\ \nonumber \\
\omega^{(3)}_{j}&=&\frac{6 \kappa _j^6-29 \kappa _j^4+26 \kappa _j^2+2 \left(2 \kappa_j^2+5\right) \kappa _j^2 \log\kappa _j^2-3}{2(16\pi)^2 \kappa _j^2 \left(\kappa _j^2-1\right){}^3}.
\end{eqnarray}
In order to compare and discuss these processes, we refer the reader to  Fig.~\ref{3evsgegraphs}.
The graphs provided there show, in logarithmic scale, contributions to both ${\rm Br}(\mu\to e\gamma)$ and ${\rm Br}(\mu\to3e)$. For each graph, both such contributions are calculated by fixing the parameters $(\kappa_1,\kappa_N)$: (2.5,2.4) for Graph (I); (3.5,3.4) for Graph (II); (7.5,7.4) for Graph (III); (10,9.9) for Graph (IV). We have used these values of the kappas because, of all the considered values in Fig~\ref{graphsforbounds}, these yield the lowest lower bounds for the $W'$ mass (for the largest possible neutrino-mixing contribution, $|v^*_{1e}v_{1\mu}|^2=1/4$) from $\mu\to3e$. In all the graphs, the solid magenta curves correspond to ${\rm Br}(\mu\to3e)$ (logarithmic scale), whereas dashed curves are associated to ${\rm Br}(\mu\to e\gamma)$ (logarithmic scale as well). We have also included two horizontal lines to each graph. The upper solid horizontal line in any graph represents the experimental upper bound on ${\rm Br}(\mu\to3e)$ from the SINDRUM Collaboration. On the other hand, the dotted horizontal line in any graph corresponds to the MEG upper bound on ${\rm Br}(\mu\to e\gamma)$, at $5.7\times10^{-13}$. The shadowed regions below each of these horizontal lines are thus the allowed regions for the corresponding decay processes. From Fig.~\ref{3evsgegraphs}, we can appreciate that the constraints on $m_{W'}$ from ${\rm Br}(\mu\to e\gamma)$ are more stringent than those from ${\rm Br}(\mu\to3e)$; in all cases, the former process pushes the lower bound on $m_{W'}$ further. This is summarized in Table~\ref{tablona},
\begin{table}[!ht]
\center
\begin{tabular}{|l|c|c|c|c|}
\hline
& Graph (I) & Graph (II) & Graph (III) & Graph (IV) 
\\ \hline
$m_{W'}^{\rm min}$, $\mu\to e\gamma$ (MEG) & 1968.75 & 1811.13 & 1055.26 & 800.28
\\ 
$m_{W'}^{\rm min}$, $\mu\to3e$ (SINDRUM) & 952.84 & 817.71 & 454.54 & 342.36
\\ 
$m_{W'}^{\rm min}$, $\mu\to3e$ (Mu3e) & 9528.35 & 8177.10 & 4545.38 & 3423.55
\\ \hline
\end{tabular}
\caption{\label{tablona} Lower bounds on $m_{W'}$ (with maximum value of neutrino-mixing dependence) for different scenarios of heavy-neutrino mass, in line with the experiments MEG and SINDRUM, and with the expected sensitivity of Mu3e.}
\end{table}
where we have also included the lower bounds on $m_{W'}$ from the expected sensitivity of Mu3e. About this, note that the expected sensitivity of Mu3e should increase bounds on the $W'$ mass by a a factor of $\sim8$.
In relation with our discussion on the mass $m_{W'}$, it is worth commenting on the analysis of Ref.~\cite{HLRZ}. That work was developed around a context that is similar to ours. The authors of this paper studied how plausible it would be to observe a right-handed heavy neutrino, of Majorana type, that couples to a $W_R$ gauge boson, also heavy, by means of the lepton-number-violating process $pp\to W_R\to l^{\pm}N\to l^{\pm}l^{\pm}jj$. Performing a thorough study of the background for this interesting process, they concluded that a $W_R$ heavy boson with mass $m_{W_R}=3\,{\rm TeV}$ or $m_{W_R}=4\,{\rm TeV}$ might be measured at the Large Hadron Collider with $90\,{\rm fb}^{-1}$ or $1\,{\rm ab}^{-1}$, respectively, if the heavy neutrino has a mass $m_{N_R}=500\,{\rm GeV}$. Translated to our language, these masses correspond to kappa parameters $\kappa_N=6$ and $\kappa_N=8$, respectively. Differently from our work, this reference pays attention to only one heavy-neutrino mass eigenstate. However, since the process explored by the authors of this reference requires that $m_{N_R}<m_{W_R}$, one may think that other right-handed neutrinos fitting such a formulation should be heavier than the $W_R$ boson, so the corresponding kappa parameters would be $<1$. \\

\subsection{New-physics contributions to $\tau\to l_\beta l_\sigma l_\sigma$}
Now we discuss the tau lepton decays. 
Lepton-flavor-violating decays of the tau lepton have been a matter of interest for the $B$ factories Belle and BABAR~\cite{Bfactories}. In 2010, the Belle and BABAR Collaborations reported that no events of tau decays $\tau\to l_\beta\,l_\sigma\,l_\sigma$ had been observed, and thus established upper bounds on the corresponding branching fractions, all of them of order $10^{-8}$~\cite{BABARLFVtau,BelleLFVtau}. Using proton-proton collision data, the LHCb Collaboration has been able to establish the upper bound $8.0\times10^{-8}$ on the branching ratio ${\rm Br}(\tau\to3\mu)$~\cite{LHCbLFVtau}. Finally, let us mention that the ATLAS Collaboration recently reported an upper bound of order $10^{-7}$ on the branching ratio of the decay $\tau\to3\mu$~\cite{ATLASLFVtau}. 
In order to estimate the branching ratios for the decay processes $\tau\to l_\beta l_\sigma l_\sigma$, we use the sets of values for $(\kappa_1,\kappa_N)$ that were considered in Fig.~\ref{3evsgegraphs}, and the lower bound on $m_{W'}$ that was imposed by the MEG constraint on ${\rm Br}(\mu\to e\gamma)$. Recall that we have been saying ``lower bound'' on $m_{W'}$ to refer to lowest possible value of this mass for the largest neutrino-mixing factor, but keep in mind that different mixing parameters would decrease this value. Let us point out that the utilization of these parameters yields branching ratios ${\rm Br}(\tau\to l_\beta l_\sigma l_\sigma)$ that automatically satisfy the experimental constraints. 
The values of the contributions are given in Tables~\ref{tau1} and \ref{tau2}. The entries of such tables are branching ratios that correspond to the kappa parameters that are indicated in the corresponding row. 
\begin{table}[!ht]
\begin{tabular}{|c|c|c|}
\hline
 & ${\rm Br}(\tau\to e\mu\mu)_{\rm max}$ & ${\rm Br}(\tau\to \mu ee)_{\rm max}$
\\ \hline
$(\kappa_1,\kappa_N)=(2.5,2.4)$ & $6.46\times10^{-15}$ & $2.24\times10^{-13}$
\\ \hline
$(\kappa_1,\kappa_N)=(3.5,3.4)$ & $4.15\times10^{-15}$ & $1.10\times10^{-13}$
\\ \hline
$(\kappa_1,\kappa_N)=(7.5,7.4)$ & $2.92\times10^{-15}$ & $5.44\times10^{-14}$
\\ \hline  
$(\kappa_1,\kappa_N)=(10,9.9)$ & $2.76\times10^{-15}$ & $4.80\times10^{-14}$
\\ \hline
\end{tabular}
\caption{\label{tau1} Values of the branching ratios for  $\tau^-\to e^-\mu^+\mu^-$ and $\tau^-\to \mu^-e^+e^-$, determined by  the parameters $\kappa_1,\kappa_N,m_{W'}$. For each selected pair $(\kappa_1,\kappa_N)$ the corresponding minimal allowed mass $m_{W'}$, given in Table~\ref{mwboundsandnumasses}, has been used.}
\end{table}
In the context of seesaw-type models, the branching ratios for the lepton-flavor-violating trilepton decays of the tau lepton into charged leptons were calculated some years ago in Ref.~\cite{IlPi}, with the conclusion that $\tau\to 3e$ and $\tau\to e^-\mu^+\mu^-$ would be the most promising options, reaching values as large as $\sim10^{-6}$. Tau decays were also considered in Ref.~\cite{GrLa}, in models of seesaw with multi-Higgs-doublets, which yielded a branching ratio ${\rm Br}(\tau\to\mu^- e^+e^-)$ around $\sim10^{-18}$. Supersymmetry with lepton flavor violation was the scenario, in Ref.~\cite{BaKo}, to explore the tau decay $\tau\to3\mu$, claiming that a value as large as $10^{-7}$ can be reached in this model. \\
\begin{table}[!ht]
\begin{tabular}{|c|c|c|}
\hline
 & ${\rm Br}(\tau\to 3e)_{\rm max}$ & ${\rm Br}(\tau\to 3\mu)_{\rm max}$
\\ \hline
$(\kappa_1,\kappa_N)=(2.5,2.4)$ & $1.08\times10^{-14}$ & $2.21\times10^{-13}$
\\ \hline
$(\kappa_1,\kappa_N)=(3.5,3.4)$ & $8.44\times10^{-15}$ & $1.05\times10^{-13}$
\\ \hline
$(\kappa_1,\kappa_N)=(7.5,7.4)$ & $7.18\times10^{-15}$ & $5.03\times10^{-14}$
\\ \hline
$(\kappa_1,\kappa_N)=(10,9.9)$ & $7.02\times10^{-15}$ & $4.38\times10^{-14}$
\\ \hline
\end{tabular}
\caption{\label{tau2} Values of the branching ratios for  $\tau^-\to e^-e^+e^-$ and $\tau^-\to \mu^-\mu^+\mu^-$, determined by  the parameters $\kappa_1,\kappa_N,m_{W'}$. For each selected pair $(\kappa_1,\kappa_N)$ the corresponding minimal allowed mass $m_{W'}$, given in Table~\ref{mwboundsandnumasses}, has been used.}
\end{table}

\section{About the mixing of light and heavy charged bosons}
\label{secmix}
In general, the existence of a heavy charged boson $W'$ opens the possibility that $W'$-$W$ mixing, involving the Standard-Model $W$ boson, occurs~\cite{SenMoother}, which we address in this subsection. Aspects about the measurement of this mixing, including $CP$-violating effects, are discussed in Ref.~\cite{BoDu}. A $W'$-$W$ mixing would be characterized by some mixing angle, $\zeta$, and a $CP$-violating phase. In the context of left-right symmetric models, the analysis of Ref.~\cite{Metal} yielded the restriction $\zeta\lesssim0.04$ on the $W'$-$W$ mixing angle, without any assumption regarding left-right symmetry nor the mass of right-handed neutrinos. A more stringent bound on this mixing has restricted $\zeta$ to range within -0.0006 to 0.0028~\cite{AFG}, but in this case {\it manifest left-right symmetry} was imposed~\cite{LaSa}. For a small mixing angle, the charged currents
\begin{equation}
\sum_{j=1}^3\sum_{\alpha=e,\mu,\tau}\left[ \zeta\frac{u_{j\alpha}}{\sqrt{2}}\,W^+_\rho\bar{N}_j\gamma^\rho P_Rl_\alpha+{\rm H.c.} \right],
\label{newCCs}
\end{equation}
which couple the Standard-Model $W$ boson and leptons to the heavy neutrinos, may be present~\cite{HLRZ,SenMoother}. We assume that the coefficients $u_{j\alpha}$ are the entries of a $3\times3$ matrix that is approximately unitary. \\

\begin{figure}[!ht]
\center
\includegraphics[width=8.6cm]{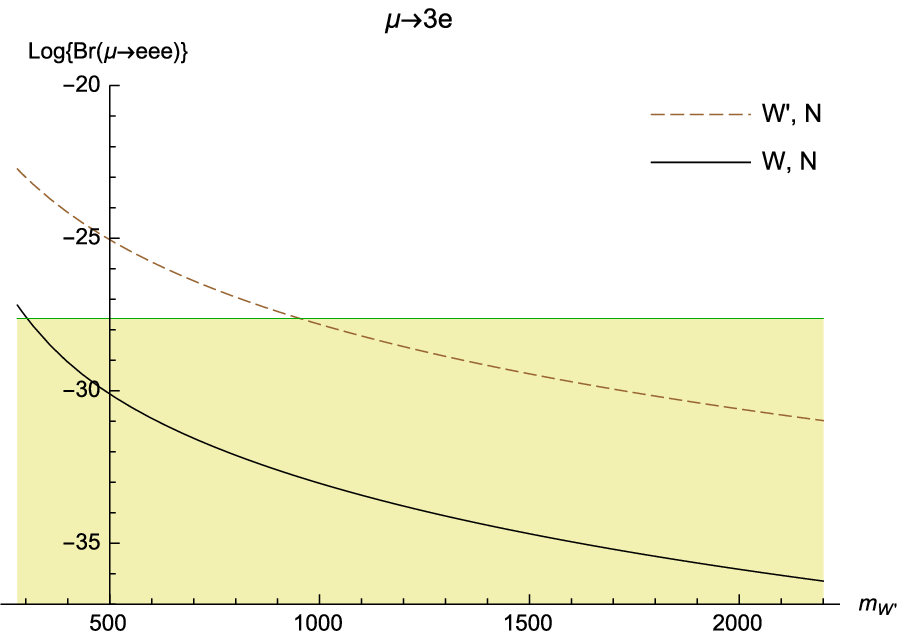}
\hspace{0.4cm}
\includegraphics[width=8.6cm]{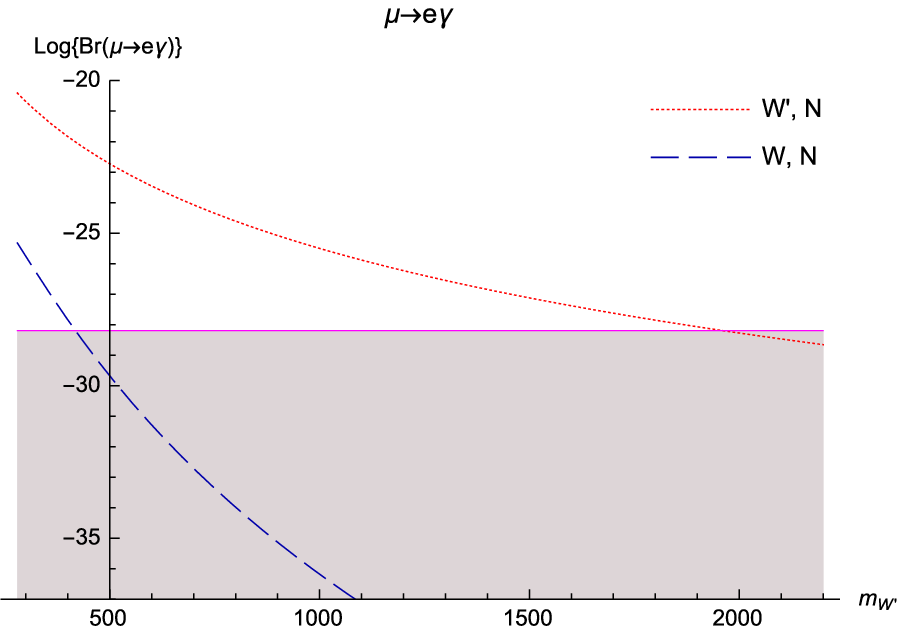}
\\
Graph (i) \hspace{8cm} Graph(ii)
\\ \vspace{0.5cm}
\includegraphics[width=8.6cm]{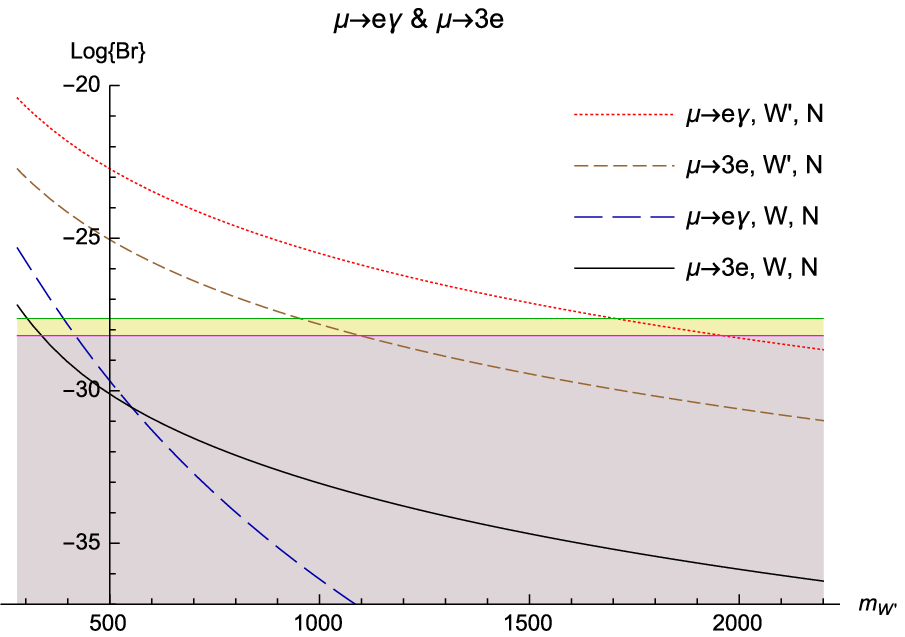}
\\
Graph (iii)
\caption{\label{compmix} }
\end{figure}

We calculated the leading $W,N$ contributions from the charged currents given in Eq.~(\ref{newCCs}) to ${\cal M}(\mu\to3e)$. We noticed that such contributions require that $\zeta\propto (\Lambda^2)^{-1}$ in order for they to decouple when $\Lambda\to\infty$; if $\zeta$ is thought of as independent of this high-energy scale, the amplitude diverges in such a limit. To this respect, let us mention Refs.~\cite{LaSa,Masso}, where, in the context of left-right symmetric models, the bound $|\zeta|\leqslant m^2_{W}/m^2_{W'}$ was derived. 
Now we take the point of view of decoupling and, for a moment, assume that this restriction on $\zeta$ holds. Thus we calculate the $W,N$ contributions to ${\rm Br}(\mu\to3e)$, which we show and compare, in Graph (i) of Fig.~\ref{compmix},  
with the $W',N$ contributions from the charged currents of Eq.~(\ref{CCs}). This and all other graphs of Fig.~\ref{compmix} were plotted in logarithmic scale, and in all cases we used the kappa parameters $\kappa_1=2.5$, $\kappa_N=2.4$. In Graph (i), we have included a solid horizontal line, which corresponds to the SINDRUM upper bound. The short-dashed plot represents the afore-calculated $W',N$ contributions to ${\rm Br}(\mu\to3e)$, from the charged currents of Eq.~(\ref{CCs}), whereas the solid curve corresponds to the $W,N$ contributions from the charged currents of Eq.~(\ref{newCCs}) to the same branching ratio, with $\zeta=m_W^2/m_{W'}^2$. Thus keep in mind that, from such a viewpoint, this plot represents an upper bound with respect to the $W'$-$W$ mixing angle. Note that the $W,N$ contribution shows the decoupling behavior that we pointed out before. 
From Graph (i), we observe that, in this decoupling scenario, the $W',N$ contributions dominate over those from $W,N$. We also compared the $W',N$ contributions with the $W,N$ contributions to the branching ratio of $\mu\to e\gamma$, for which we provide Graph (ii) of Fig.~\ref{compmix}. In this graph, the dotted plot is the $W',N$ contribution, originated in the charged currents given in Eq.~(\ref{CCs}), to ${\rm Br}(\mu\to e\gamma)$, while the long-dashed curve corresponds to the $W,N$ contributions to this branching ratio, produced by the charged currents of Eq.~(\ref{newCCs}). The solid horizontal line represents the MEG bound on ${\rm Br}(\mu\to e\gamma)$. Again, with the criterion of decoupling, the $W',N$ contributions dominate over those from $W,N$, which can be appreciated in this graph. Finally, we have included Graph (iii) in Fig.~\ref{compmix} to put all contributions from Graphs (i) and (ii) together. This graph illustrates that the largest decoupling contributions are those from $W',N$ to ${\rm Br}(\mu\to e\gamma)$, for $300\,{\rm GeV}\leqslant m_{W'}$. Graph (iii) also shows that the $W,N$ contributions to ${\rm Br}(\mu\to e\gamma)$ are larger than the contributions to ${\rm Br}(\mu\to3e)$, from the same charged currents (Eq.~(\ref{newCCs})), for $300\,{\rm GeV}\leqslant m_{W'}\lesssim552\,{\rm GeV}$. If $m_{W'}\gtrsim552\,{\rm GeV}$, this situation is reversed, that is, ${\rm Br}(\mu\to e\gamma)<{\rm Br}(\mu\to3e)$ in the case of the $W,N$ contributions. This change takes place because the $W,N$ contributions to ${\rm Br}(\mu\to e\gamma)$ decrease rapidly, which, in turn, is a consequence of the fact that these contributions, contrastingly to those to ${\rm Br}(\mu\to3e)$, are decoupling even if $\zeta$ is just a parameter, completely independent of  the high-energy scale $\Lambda$. To this respect, note that both the leading $W,N$ contributions to ${\cal M}(\mu\to3e)$ and those corresponding to ${\cal M}(\mu\to e\gamma)$ are determined by the same one-loop vertex $e\mu\gamma$. In the case of $\mu\to e\gamma$, this vertex contributes on shell, while the decay process $\mu\to3e$ receives off-shell contributions from this vertex. We have verified that the nondecoupling effects of the leading contributions to ${\cal M}(\mu\to3e)$ lie within those terms of the vertex $e\mu\gamma$ that vanish in the on-shell case. \\

\begin{figure}[!ht]
\center
\includegraphics[width=8.6cm]{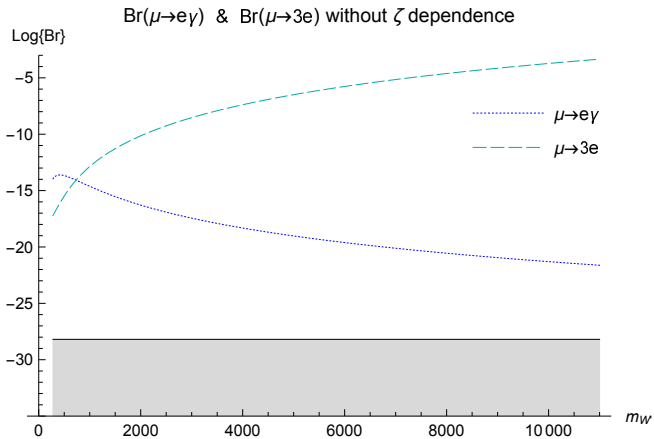}
\caption{\label{nomixdep} High-energy behavior of ${\rm Br}(\mu\to e\gamma)$ and ${\rm Br}(\mu\to3e)$ without dependence on the mixing angle $\zeta$. The plots are given in logarithmic scale. The shadowed region is allowed by the SINDRUM bound.}
\end{figure}

\begin{figure}[!ht]
\center
\includegraphics[width=15cm]{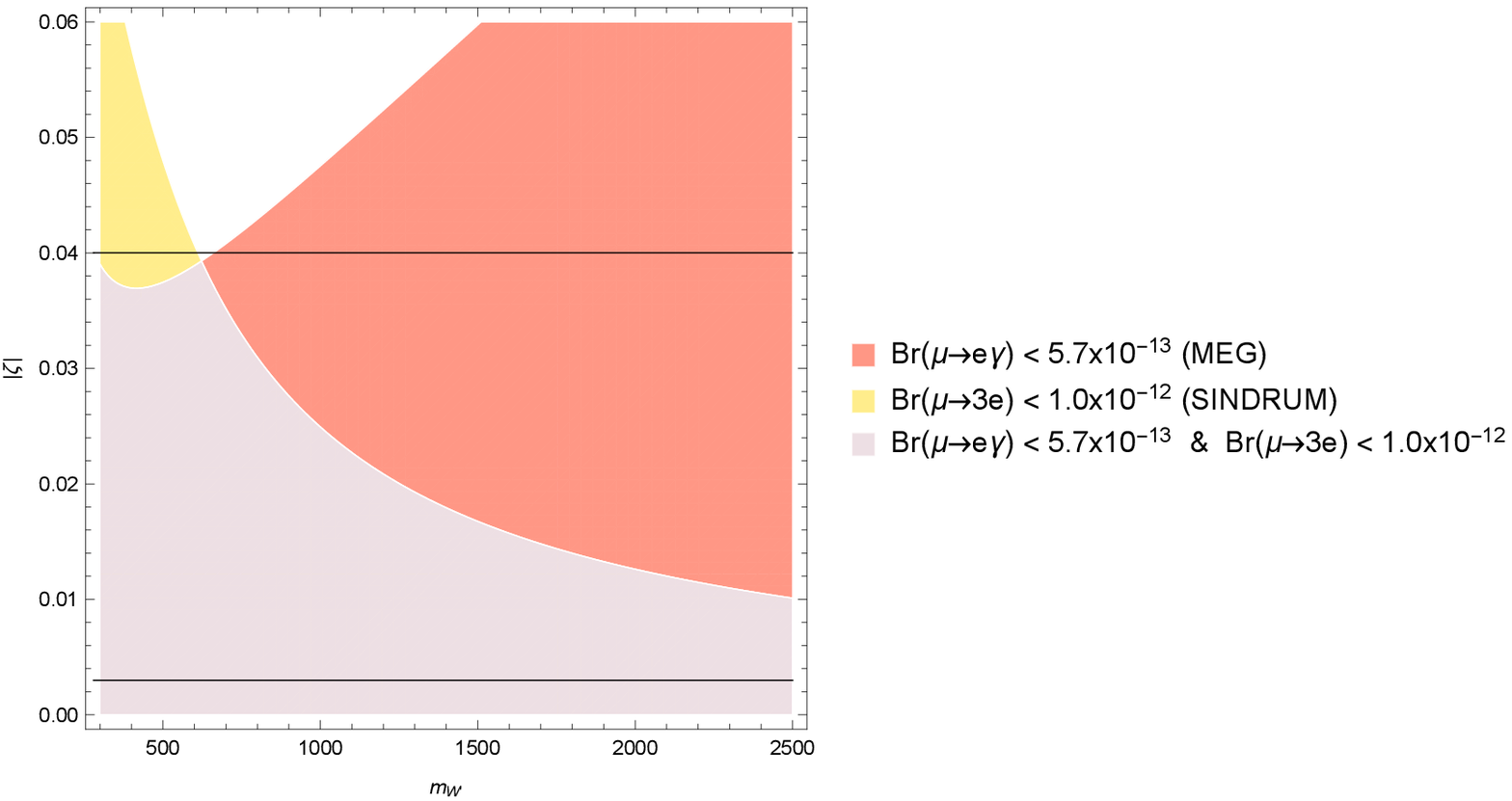}
\includegraphics[width=15cm]{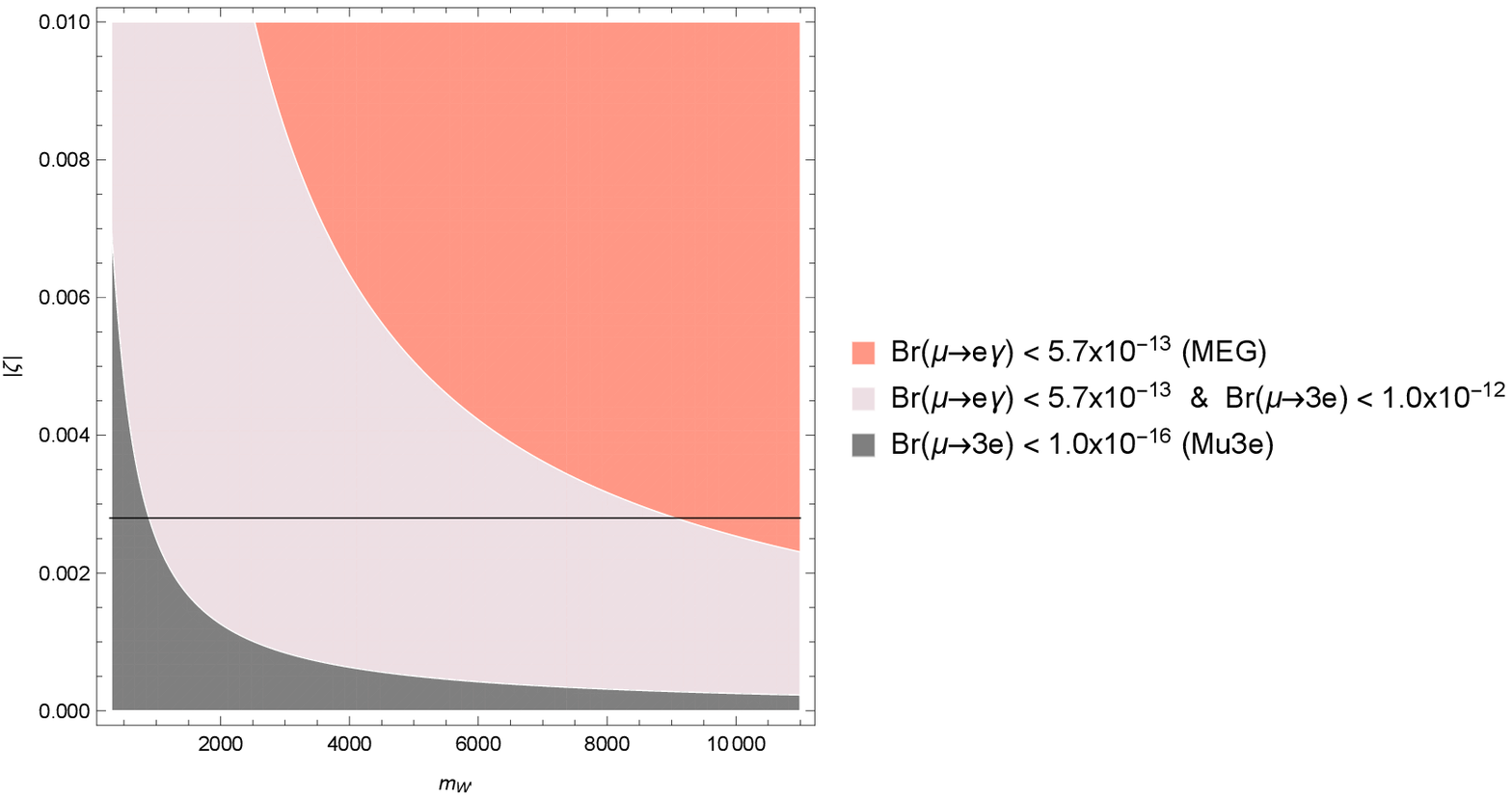}
\caption{\label{graphsmix} Allowed regions in the $(|\zeta|,m_{W'})$ plane, as stated by the limits on $\mu\to e\gamma$ (MEG) and $\mu\to3e$ (SINDRUM). The colorless region, which only appears in the upper graph, would be already discarded by these bounds. The lower graph provides a region that would be still allowed by the expected sensitivity of Mu3e (gray).}
\end{figure}

For the following discussion we consider the mixing angle $\zeta$, rather, as a parameter. Under such circumstances, we aim at bounding the $W'$-$W$ mixing. Using the kappa parameters $\kappa_1=2.5$, $\kappa_N=2.4$, and omitting the mixing-angle dependence, we have plotted the graph of Fig.~\ref{nomixdep},
which shows, within such a context, the behavior of the $W,N$ contributions to ${\rm Br}(\mu\to e\gamma)$ (dotted curve) and ${\rm Br}(\mu\to3e)$ (dashed curve) for $W'$-mass values ranging from $300\,{\rm GeV}$ to $11\,{\rm TeV}$, both in logarithmic scale. The solid horizontal line in this figure represents the SINDRUM upper bound on ${\rm Br}(\mu\to3e)$. This graph illustrates our discussion of the last paragraph: the $W,N$ contributions to ${\rm Br}(\mu\to e\gamma)$ decouple independently of whether $\zeta$ is restricted by the high-energy scale $\Lambda$ or not, but the $W,N$ contributions to ${\rm Br}(\mu\to3e)$ without the assumption that $\zeta\leqslant m^2_W/m^2_{W'}$ are non-decoupling. This observation is important because it determines that the most stringent restrictions on the mixing angle $\zeta$ shall be set by ${\rm Br}(\mu\to3e)$. We provide Fig.~\ref{graphsmix}, 
whose graphs, defined in the parameter space $(|\zeta|,m_{W'})$, were made with kappa parameters $\kappa_1=2.5$, $\kappa_N=2.4$. The upper graph of this figure, where $|\zeta|$ and $m_{W'}$ range within $300\,{\rm GeV}\leqslant m_{W'}\leqslant2.5\,{\rm TeV}$ and $0\leqslant|\zeta|\leqslant 0.06$, exhibits a region in light red, which is allowed by the condition ${\rm Br}(\mu\to e\gamma)<5.7\times10^{-13}$, in accordance with the MEG bound. Another region, colored in yellow, corresponds to the set of values that satisfy the condition ${\rm Br}(\mu\to3e)<1.0\times10^{-12}$, from the SINDRUM Collaboration. Both of these regions extend downwards and rightwards, and they share points, defining the third, and most important, region, which appears in pink: the intersection of the allowed regions for both processes, that is, the set of points in the parameter space $(|\zeta|,m_{W'})$ that is simultaneously allowed by the experimental restrictions on these physical processes. For comparison purposes, we have also added two horizontal solid lines. The upper line corresponds to the value $|\zeta|=0.04$, which is the upper limit given by Ref.~\cite{Metal}; the lower horizontal line, on the other hand, is $|\zeta|=0.0028$, in accordance with the upper bound of Ref.~\cite{AFG}. There is also a colorless region, which is the one already ruled out by experimental data. For the SINDRUM and MEG constraints to be fulfilled, we observe that the mixing angle can be as large as $\zeta\approx0.039$, but this would require a $W'$ mass as small as $m_{W'}\approx620\,{\rm GeV}$. Evidently, larger values of the $W'$ mass constrain the mixing angle $\zeta$ further. The lower graph of Fig.~\ref{graphsmix} involves a wider range of $m_{W'}$ values and smaller values of $|\zeta|$: $300\,{\rm GeV}\leqslant m_{W'}\leqslant11\,{\rm TeV}$ and $0\leqslant |\zeta|\leqslant0.01$. Moreover, this graph includes a small region of small-$\zeta$ values, colored in gray, which corresponds to points in the parameter space $(|\zeta|,m_{W'})$ that would still be allowed by the expected sensitivity of the Mu3e experiment, in case of no detection of $\mu\to 3e$. Such a sensitivity would impose stringent constraints; for instance, the lower bound $|\zeta|<0.0028$, provided by Ref.~\cite{AFG} and represented in this graph by the horizontal solid line, would already be the upper bound from Mu3e at $m_{W'}\approx891\,{\rm GeV}$, but at $m_{W'}\approx9070\,{\rm GeV}$, which would be the upper bound for the combined restrictions from SINDRUM and MEG, the Mu3e experiment would restrict the $W'$-$W$ mixing angle as stringently as $|\zeta|\lesssim2.7\times10^{-4}$.


\section{Conclusions}
\label{conc}
In the present paper, we have calculated, at the one-loop level, contributions from charged currents featuring heavy neutrinos and a heavy charged gauge boson to decays of the muon and tau lepton into three charged leptons $l_\alpha\to l_\beta\,l_\sigma\,l_\sigma$, in which lepton flavor is not preserved. Such effects are possible due to neutrino mixing, which does not occur in the Standard Model, but certainly is a phenomenon of nature, according to the experimental evidence of neutrino oscillations. We calculated the dominant contributions to the amplitudes of these decays, which come from reducible diagrams in which loop subdiagrams are linked to an electromagnetic vertex by means of a photon propagator; this induces an enhancement of contributions by a propagator-pole effect. Moreover, we have emphasized that our results do not distinguish among Dirac and Majorana neutrinos, because the Feynman diagrams that produce the leading contributions are the same in both cases. We have proven explicitly that our results are ultraviolet-finite and gauge invariant, with respect to the electromagnetic group, and we have shown that they decouple. We considered a mass spectrum for the heavy neutrinos in which two neutrinos have quasi-degenerate masses, but we left the mass of the third neutrino different. Then we calculated the branching fraction ${\rm Br}(l_\alpha\to l_\beta\,l_\sigma\,l_\sigma)$ and implemented it to specific lepton decays. We first considered the decay $\mu\to3e$, which we used to establish sets of parameters that are in agreement with the currently most stringent upper bounds on such a branching ratio, set by the SINDRUM Collaboration. Such parameters were the $W'$ mass and two parameters, $\kappa_1$ and $\kappa_N$, which define, for a large high-energy symmetry-breaking scale, a simple relation among $m_{W'}$ and the heavy-neutrino masses $m_j$: $m_{W'}\approx\kappa_j m_j$, for any $j$. Using the SINDRUM limits we established, for different sets of kappa parameters $\kappa_j$, lower bounds on the $W'$ mass. These bounds were set for the maximum value of the neutrino-mixing dependence, although we pointed out that, in general, different sets of neutrino-mixing parameters should attenuate the contributions to ${\rm Br}(\mu\to3e)$, thus relaxing lower bounds on the $W'$ mass. The impact of neutrino mixing on ${\rm Br}(\mu\to3e)$ was then investigated, for which allowed regions, consistent with the limits from SINDRUM, were provided. The expected sensitivity of the Mu3e experiment was also taken into account in the discussion on the effects of neutrino mixing. We also considered the decay process $\mu\to e\gamma$, whose branching ratio compared with ${\rm Br}(\mu\to3e)$. Using the limits on ${\rm Br}(\mu\to e\gamma)$, reported by the MEG Collaboration, we determined that the limits on the $W'$ set by this process are more stringent than those obtained from ${\rm Br}(\mu\to3e)$. We then used sets of parameters, established by means of $\mu\to e\gamma$ and the MEG bound, to determine the size of the contributions to the branching ratios of the tau decays $\tau\to l_\beta\,l_\sigma\,l_\sigma$, finding that they are $\sim10^{-15}-10^{-13}$. These contributions are far below the upper bounds reported by the Belle and BABAR Collaborations. A possible mixing of $W'$ with the Standard-Model $W$ boson was explored, for which we studied the impact of the mixing angle $\zeta$, that parametrizes such a mixing, on charged currents that couple the $W$ boson and Standard-Model leptons to the set of heavy neutrinos. In the context of this discussion, we showed regions in the parameter space $(|\zeta|,m_{W'})$ for which the processes $\mu\to e\gamma$ and $\mu\to3e$ remain allowed by the experimental data from MEG and SINDRUM, though the expected sensitivity if Mu3e was considered as well. While the combined restrictions form these experiments allow a mixing angle as large as $|\zeta|\sim0.04$, for a relatively light $W'$ boson, the upper limit on $\zeta$ decreases as the $W'$ mass increases. Moreover, we found that the expected sensitivity of the Mu3e experiment would establish stringent bounds on $\zeta$, which thus could be as large as $\zeta\sim10^{-4}$ for a $W'$ boson with mass in the range of few TeVs.

\begin{acknowledgements}
The authors acknowledge financial support from CONACYT (M\'exico). In addition, H.N.S. and J.J.T. acknowledge financial support from Sistema Nacional de Investigadores (SNI) (M\'exico). H.N.S. also acknowledges financial support from Programa para el Desarrollo Profesional Docente, para el Tipo Superior (PRODEP) (M\'exico), Project No. DSA/103.5/16/10420.
\end{acknowledgements}

\appendix

\section{Explicit expressions of electromagnetic form factors}
\label{appfs}
In this Appendix we exhibit the explicit expressions for the factors $f^j_{\Omega,0}$, $f^j_{\Omega,12}$, $f^j_{\Omega,23}$, $f^j_{\Omega,34}$, and $f^j_{\Omega,5}$, which constitute the ultraviolet-free factors $f^j_\Omega$, given in Eq.~(\ref{finfs}):
\begin{equation}
\Delta_Q=32 \pi ^2 m_{\alpha } m_{\beta } \left(m_{\beta }-m_{\alpha }\right) m_{W'}^2 \left(\left(m_{\alpha }-m_{\beta
   }\right){}^2-q^2\right) \left(\left(m_{\alpha }+m_{\beta }\right){}^2-q^2\right){}^2,
\end{equation}
\begin{equation}
\Delta_A=32 \pi ^2 m_{\alpha } m_{\beta } \left(m_{\alpha }+m_{\beta }\right) m_{W'}^2 \left(\left(m_{\alpha }-m_{\beta
   }\right){}^2-q^2\right){}^2 \left(\left(m_{\alpha }+m_{\beta }\right){}^2-q^2\right){}^2,
\end{equation}
\begin{equation}
\Delta_M=32 \pi ^2 m_{\alpha } m_{\beta } m_{W'}^2 \left(\left(m_{\alpha }-m_{\beta
   }\right){}^2-q^2\right) \left(\left(m_{\alpha }+m_{\beta }\right){}^2-q^2\right){}^2,
\end{equation}
\begin{equation}
\Delta_E=32 i \pi ^2 m_{\alpha } m_{\beta } m_{W'}^2 \left(\left(m_{\alpha }-m_{\beta }\right){}^2-q^2\right){}^2
   \left(\left(m_{\alpha }+m_{\beta }\right){}^2-q^2\right),
\end{equation}
\begin{eqnarray}
f^j_{Q,0}&=&-q^2 m_{\alpha } m_{\beta } \left(m_{\alpha }-m_{\beta }\right) \left(m_{\alpha }^4-2 m_{\alpha }^2 \left(m_{\beta
   }^2+q^2\right)+\left(m_{\beta }^2-q^2\right){}^2\right) \left(m_j^2+m_{\alpha } m_{\beta }+2 m_{W'}^2\right),
\end{eqnarray}
\begin{equation}
f^j_{Q,12}=q^2 \left(m_{\alpha }-m_{\beta }\right) \left(m_j-m_{W'}\right) \left(m_j+m_{W'}\right) \left(m_{\alpha }^4-2 m_{\alpha }^2\left(m_{\beta }^2+q^2\right)+\left(m_{\beta }^2-q^2\right){}^2\right) \left(m_j^2+m_{\alpha } m_{\beta }+2 m_{W'}^2\right),
\end{equation}
\begin{eqnarray}
f^j_{Q,23}&=&-q^2 \left(m_{\alpha }-m_{\beta }\right) \big(m_j^4 \left(-\left(m_{\alpha }+m_{\beta }\right){}^2 \left(-6 m_{\alpha }m_{\beta }+m_{\alpha }^2+m_{\beta }^2\right)+2 q^2 \left(m_{\alpha } m_{\beta }+m_{\alpha }^2+m_{\beta}^2\right)-(q^2)^2\right)
\nonumber \\ &&
-m_j^2 \big(m_{W'}^2 \left(\left(m_{\alpha }+m_{\beta }\right){}^2 \left(-6 m_{\alpha } m_{\beta}+m_{\alpha }^2+m_{\beta }^2\right)-2 q^2 \left(m_{\alpha } m_{\beta }+m_{\alpha }^2+m_{\beta}^2\right)+(q^2)^2\right)+m_{\alpha } m_{\beta } \left(m_{\alpha }+m_{\beta }\right){}^2
\nonumber \\ &&
\times
\left(-2 m_{\alpha } m_{\beta }+5m_{\alpha }^2+5 m_{\beta }^2-5 q^2\right)\big)+m_{\alpha }^2 m_{\beta }^2 \left(4 m_{\alpha } m_{\beta } \left(m_{\alpha}+m_{\beta }\right){}^2-q^2 \left(m_{\alpha }^2+m_{\beta }^2\right)+(q^2)^2\right)
+2 m_{W'}^4
\nonumber \\ &&
\times
\left(\left(m_{\alpha }+m_{\beta}\right){}^2 \left(-6 m_{\alpha } m_{\beta }+m_{\alpha }^2+m_{\beta }^2\right)-2 q^2 \left(m_{\alpha } m_{\beta }+m_{\alpha}^2+m_{\beta }^2\right)+(q^2)^2\right)+m_{\alpha } m_{\beta } m_{W'}^2 \big(\left(m_{\alpha }+m_{\beta }\right){}^4
\nonumber \\ &&
+2 q^2\left(7 m_{\alpha } m_{\beta }+2 m_{\alpha }^2+2 m_{\beta }^2\right)-5 (q^2)^2\big)\big),
\end{eqnarray}
\begin{eqnarray}
f^j_{Q,34}&=&q^2 m_{\alpha } \big(m_j^4 \left(3 m_{\alpha }^3 m_{\beta }+m_{\alpha }^4+8 m_{\beta }^4-m_{\alpha }^2 \left(m_{\beta }^2+2q^2\right)+m_{\alpha } \left(5 m_{\beta }^3-3 q^2 m_{\beta }\right)-3 q^2 m_{\beta }^2+(q^2)^2\right)+m_j^2
\nonumber \\ &&
\times
\big(m_{W'}^2 \big(3m_{\alpha }^3 m_{\beta }
+m_{\alpha }^4+8 m_{\beta }^4-m_{\alpha }^2 \left(m_{\beta }^2+2 q^2\right)+m_{\alpha } \left(5m_{\beta }^3-3 q^2 m_{\beta }\right)-3 q^2 m_{\beta }^2+(q^2)^2\big)-m_{\beta } \left(m_{\alpha }+m_{\beta }\right)
\nonumber \\ &&
\times
\big(-m_{\alpha }^3 m_{\beta }+m_{\alpha }^4+6 m_{\beta }^4
+m_{\alpha }^2 \left(13 m_{\beta }^2-2 q^2\right)+m_{\alpha }m_{\beta } \left(q^2-3 m_{\beta }^2\right)-7 q^2 m_{\beta }^2+(q^2)^2\big)\big)
+m_{\alpha } m_{\beta }^3
\nonumber \\ &&
\times
\big(m_{\alpha }\left(m_{\alpha }+m_{\beta }\right) \left(-2 m_{\alpha } m_{\beta }+3 m_{\alpha }^2+7 m_{\beta }^2\right)
-q^2 \left(3m_{\alpha } m_{\beta }+4 m_{\alpha }^2+m_{\beta }^2\right)+(q^2)^2\big)-2 m_{W'}^4 \big(3 m_{\alpha }^3 m_{\beta}
\nonumber \\ &&
+m_{\alpha }^4+8 m_{\beta }^4-m_{\alpha }^2 \left(m_{\beta }^2+2 q^2\right)+m_{\alpha } \left(5 m_{\beta }^3-3 q^2 m_{\beta}\right)
-3 q^2 m_{\beta }^2+(q^2)^2\big)+m_{\beta } m_{W'}^2 \big(7 m_{\alpha }^4 m_{\beta }+3 m_{\alpha }^5
\nonumber \\ &&
+m_{\alpha }\left(10 m_{\beta }^4+5 q^2 m_{\beta }^2+3 (q^2)^2\right)-m_{\alpha }^3 \left(5 m_{\beta }^2+6 q^2\right)-m_{\alpha }^2 \big(3m_{\beta }^3
+5 q^2 m_{\beta }\big)
\nonumber \\ &&
+2 m_{\beta } \left(m_{\beta }^2-q^2\right) \left(2 m_{\beta }^2+q^2\right)\big)\big),
\end{eqnarray}
\begin{eqnarray}
f^j_{Q,5}&=&-2 q^2 m_{\alpha } m_{\beta } \left(m_{\alpha }-m_{\beta }\right) \big(m_j^4 \left(-3 \left(m_{\alpha }+m_{\beta }\right){}^2\left(m_{\alpha }^2+m_{\beta }^2\right)+q^2 \left(7 m_{\alpha } m_{\beta }+2 m_{\alpha }^2+2 m_{\beta}^2\right)+(q^2)^2\right)+m_j^2
\nonumber \\ &&
\times
\big((q^2)^2 \left(3 m_{\alpha } m_{\beta }+m_{\alpha }^2+m_{\beta }^2-2 m_{W'}^2\right)-q^2\big(5 m_{\alpha }^3 m_{\beta }+3 m_{\alpha }^2 m_{\beta }^2+5 m_{\alpha } m_{\beta }^3+2 m_{\alpha }^4+2 m_{\beta }^4-4m_{\alpha } m_{\beta } m_{W'}^2
\nonumber \\ &&
+3 m_{W'}^4\big)+\left(m_{\alpha }+m_{\beta }\right){}^2 \left(4 m_{\alpha }^2 m_{\beta}^2+m_{\alpha }^4+m_{\beta }^4+2 \left(m_{\alpha }^2+m_{\beta }^2\right) m_{W'}^2-6 m_{W'}^4\right)\big)+m_j^6 \left(2\left(m_{\alpha }+m_{\beta }\right){}^2+q^2\right)
\nonumber \\ &&
+m_{W'}^4 \left(2 \left(m_{\alpha }+m_{\beta }\right){}^2 \left(-3m_{\alpha } m_{\beta }+2 m_{\alpha }^2+2 m_{\beta }^2\right)-q^2 \left(11 m_{\alpha } m_{\beta }+8 m_{\alpha }^2+8 m_{\beta}^2\right)+4 (q^2)^2\right)+m_{\alpha }^2 m_{\beta }^2 
\nonumber \\ &&
\times
\big(q^2 \big(3 m_{\alpha } m_{\beta }
+m_{\alpha }^2+m_{\beta}^2\big)-\left(m_{\alpha }+m_{\beta }\right){}^2 \left(m_{\alpha }^2+m_{\beta }^2\right)\big)+2 m_{W'}^6 \left(2\left(m_{\alpha }+m_{\beta }\right){}^2+q^2\right)
\nonumber \\ &&
-2 m_{W'}^2 \left(\left(m_{\alpha }+m_{\beta }\right){}^2-q^2\right)\big(m_{\alpha }^4
+m_{\beta }^4-q^2 \left(m_{\alpha }^2+m_{\beta }^2\right)\big)\big),
\end{eqnarray}
\begin{equation}
f^j_{A,0}=m_{\alpha } m_{\beta } \left(m_{\alpha }+m_{\beta }\right) \left(\left(m_{\alpha
   }-m_{\beta }\right){}^2-q^2\right) \left(\left(m_{\alpha }+m_{\beta
   }\right){}^2-q^2\right){}^2 \left(-m_j^2+m_{\alpha } m_{\beta }-2 m_{W'}^2\right),
\end{equation}
\begin{equation}
f^j_{A,12}=-\left(m_{\alpha }+m_{\beta }\right) \left(m_j-m_{W'}\right) \left(m_j+m_{W'}\right) \left(\left(m_{\alpha }-m_{\beta
   }\right){}^2-q^2\right) \left(\left(m_{\alpha }+m_{\beta }\right){}^2-q^2\right){}^2 \left(m_j^2-m_{\alpha } m_{\beta
   }+2 m_{W'}^2\right),
\end{equation}
\begin{eqnarray}
f^j_{A,23}&=&-\left(m_{\alpha }+m_{\beta }\right) \left(\left(m_{\alpha }+m_{\beta}\right){}^2-q^2\right) \big(m_j^4 \big(\left(m_{\alpha }-m_{\beta }\right){}^2 \left(6m_{\alpha } m_{\beta }+m_{\alpha }^2+m_{\beta }^2\right)-2 q^2 \left(-m_{\alpha }m_{\beta }+m_{\alpha }^2+m_{\beta }^2\right)
\nonumber \\ &&
+(q^2)^2\big)+m_j^2 \big(m_{W'}^2\left(\left(m_{\alpha }-m_{\beta }\right){}^2 \left(6 m_{\alpha }m_{\beta }+m_{\alpha}^2+m_{\beta }^2\right)-2 q^2 \left(-m_{\alpha } m_{\beta }+m_{\alpha }^2+m_{\beta}^2\right)+(q^2)^2\right)-m_{\alpha } m_{\beta }
\nonumber \\ &&
\times
\left(m_{\alpha }-m_{\beta }\right){}^2
\left(2 m_{\alpha } m_{\beta }+5 m_{\alpha }^2+5 m_{\beta }^2-5q^2\right)\big)+m_{\alpha }^2 m_{\beta }^2 \left(4 m_{\alpha } m_{\beta }\left(m_{\alpha }-m_{\beta }\right){}^2+q^2 \left(m_{\alpha }^2+m_{\beta}^2\right)-(q^2)^2\right)
\nonumber \\ &&
-2 m_{W'}^4 \left(\left(m_{\alpha }-m_{\beta }\right){}^2 \left(6m_{\alpha } m_{\beta }+m_{\alpha }^2+m_{\beta }^2\right)-2 q^2 \left(-m_{\alpha }m_{\beta }+m_{\alpha }^2+m_{\beta }^2\right)+(q^2)^2\right)
\nonumber \\ &&
+m_{\alpha } m_{\beta } m_{W'}^2\big(\left(m_{\alpha }-m_{\beta }\right){}^4+2 q^2 \left(-7 m_{\alpha } m_{\beta }+2m_{\alpha }^2+2 m_{\beta }^2\right)-5 (q^2)^2\big)\big),
\end{eqnarray}
\begin{eqnarray}
f^j_{A,34}&=&-m_{\alpha } \left(\left(m_{\alpha }+m_{\beta }\right){}^2-q^2\right) \big(m_j^4 \big(-3m_{\alpha }^3 m_{\beta }+m_{\alpha }^4+8 m_{\beta }^4-m_{\alpha }^2 \left(m_{\beta}^2+2 q^2\right)+m_{\alpha } \left(3 q^2 m_{\beta }-5 m_{\beta }^3\right)-3 q^2 m_{\beta}^2
\nonumber \\ &&
+(q^2)^2\big)
+m_j^2 \big(m_{\beta } \left(m_{\alpha }-m_{\beta }\right)\big(m_{\alpha }^3 m_{\beta }+13 m_{\alpha }^2 m_{\beta }^2+3 m_{\alpha } m_{\beta}^3+m_{\alpha }^4+6 m_{\beta }^4-q^2 \left(m_{\alpha } m_{\beta }+2 m_{\alpha }^2+7m_{\beta }^2\right)
\nonumber \\ &&
+(q^2)^2\big)
+m_{W'}^2 \left(-3 m_{\alpha }^3 m_{\beta }+m_{\alpha}^4+8 m_{\beta }^4-m_{\alpha }^2 \left(m_{\beta }^2+2 q^2\right)+m_{\alpha } \left(3 q^2m_{\beta }-5 m_{\beta }^3\right)-3 q^2 m_{\beta }^2+(q^2)^2\right)\big)
\nonumber \\ &&
-m_{\alpha }m_{\beta }^3 \big(m_{\alpha } \big(m_{\alpha }
-m_{\beta }\big) \left(2 m_{\alpha }m_{\beta }+3 m_{\alpha }^2+7 m_{\beta }^2\right)-q^2 \left(-3 m_{\alpha } m_{\beta }+4m_{\alpha }^2+m_{\beta }^2\right)+(q^2)^2\big)-2 m_{W'}^4 
\nonumber \\ &&
\times
\big(-3 m_{\alpha }^3 m_{\beta}+m_{\alpha }^4+8 m_{\beta }^4
-m_{\alpha }^2 \left(m_{\beta }^2+2 q^2\right)+m_{\alpha }\left(3 q^2 m_{\beta }-5 m_{\beta }^3\right)-3 q^2 m_{\beta }^2+(q^2)^2\big)-m_{\beta }m_{W'}^2 
\nonumber \\ &&
\times
\big(-7 m_{\alpha }^4 m_{\beta }+3 m_{\alpha }^5+m_{\alpha } \big(10m_{\beta }^4+5 q^2 m_{\beta }^2+3 (q^2)^2\big)+2 m_{\beta } \left(-2 m_{\beta }^4+q^2m_{\beta }^2+(q^2)^2\right)-m_{\alpha }^3 \left(5 m_{\beta }^2+6 q^2\right)
\nonumber \\ &&
+m_{\alpha }^2\left(3 m_{\beta }^3+5 q^2 m_{\beta }\right)\big)\big),
\end{eqnarray}
\begin{eqnarray}
f^j_{A,5}&=&2 m_{\alpha } m_{\beta } \left(m_{\alpha }+m_{\beta }\right) \big(m_j^6 \left(-2\left(m_{\alpha }^2-m_{\beta }^2\right){}^2+q^2 \left(-6 m_{\alpha } m_{\beta }+m_{\alpha}^2+m_{\beta }^2\right)+(q^2)^2\right)+m_j^4 \left(\left(m_{\alpha }+m_{\beta}\right){}^2-q^2\right)
\nonumber \\ &&
\times
\left(3 \left(m_{\alpha }-m_{\beta }\right){}^2 \left(m_{\alpha}^2+m_{\beta }^2\right)+q^2 \left(7 m_{\alpha } m_{\beta }-2 m_{\alpha }^2-2 m_{\beta}^2\right)-(q^2)^2\right)-m_j^2 \left(\left(m_{\alpha }+m_{\beta }\right){}^2-q^2\right)
\nonumber \\ &&
\times
\big((q^2)^2 \big(-3 m_{\alpha } m_{\beta }+m_{\alpha }^2+m_{\beta }^2-2m_{W'}^2\big)-q^2 \big(-5 m_{\alpha }^3 m_{\beta }+3 m_{\alpha }^2 m_{\beta }^2-5m_{\alpha } m_{\beta }^3+2 m_{\alpha }^4+2 m_{\beta }^4+4 m_{\alpha } m_{\beta }m_{W'}^2
\nonumber \\ &&
+3 m_{W'}^4\big)+\big(m_{\alpha }
-m_{\beta }\big){}^2 \left(4 m_{\alpha}^2 m_{\beta }^2+m_{\alpha }^4+m_{\beta }^4+2 \left(m_{\alpha }^2+m_{\beta }^2\right)m_{W'}^2-6 m_{W'}^4\right)\big)+\left(\left(m_{\alpha }+m_{\beta}\right){}^2-q^2\right)
\nonumber \\ &&
\times
\big(m_{W'}^4 \big(-2 \left(m_{\alpha }-m_{\beta }\right){}^2\left(3 m_{\alpha } m_{\beta }+2 m_{\alpha }^2+2 m_{\beta }^2\right)+q^2 \left(-11m_{\alpha } m_{\beta }+8 m_{\alpha }^2+8 m_{\beta }^2\right)-4 (q^2)^2\big)
\nonumber\\&& 
+m_{\alpha }^2m_{\beta }^2 \big(\left(m_{\alpha }-m_{\beta }\right){}^2 \left(m_{\alpha }^2+m_{\beta}^2\right)
-q^2 \left(-3 m_{\alpha } m_{\beta }+m_{\alpha }^2+m_{\beta }^2\right)\big)+2m_{W'}^6 \left(-2 \left(m_{\alpha }-m_{\beta }\right){}^2-q^2\right)
\nonumber \\ &&
+2 m_{W'}^2\left(\left(m_{\alpha }-m_{\beta }\right){}^2-q^2\right) \big(m_{\alpha }^4+m_{\beta}^4
-q^2 \left(m_{\alpha }^2+m_{\beta }^2\right)\big)\big)\big),
\end{eqnarray}
\begin{equation}
f^j_{M,0}=m_{\alpha } m_{\beta } \left(m_{\alpha }+m_{\beta }\right) \left(m_{\alpha }^4-2 m_{\alpha
   }^2 \left(m_{\beta }^2+q^2\right)+\left(m_{\beta }^2-q^2\right){}^2\right)
   \left(m_j^2+m_{\alpha } m_{\beta }+2 m_{W'}^2\right),
\end{equation}
\begin{equation}
f^j_{M,12}=-\left(m_{\alpha }+m_{\beta }\right) \left(m_j-m_{W'}\right) \left(m_j+m_{W'}\right)
   \left(m_{\alpha }^4-2 m_{\alpha }^2 \left(m_{\beta }^2+q^2\right)+\left(m_{\beta
   }^2-q^2\right){}^2\right) \left(m_j^2+m_{\alpha } m_{\beta }+2 m_{W'}^2\right),
\end{equation}
\begin{eqnarray}
f^j_{M,23}&=&-\left(m_{\alpha }+m_{\beta }\right) \big(m_j^2 \big((q^2)^2 \left(m_{W'}^2-4 m_{\alpha }m_{\beta }\right)-q^2 \big(m_{\alpha } m_{\beta } \left(2 m_{\alpha } m_{\beta }-3m_{\alpha }^2-3 m_{\beta }^2\right)+2 \big(3 m_{\alpha } m_{\beta }
+m_{\alpha}^2
\nonumber \\ &&
+m_{\beta }^2\big) m_{W'}^2\big)+\left(m_{\alpha }^2-m_{\beta }^2\right){}^2\left(m_{\alpha } m_{\beta }+m_{W'}^2\right)\big)+m_j^4 \left(m_{\alpha }^4-2m_{\alpha }^2 \left(m_{\beta }^2+q^2\right)-6 q^2 m_{\alpha } m_{\beta }+\left(m_{\beta}^2-q^2\right){}^2\right)
\nonumber \\ &&
-2 m_{W'}^4 \left(\left(m_{\alpha }^2-m_{\beta }^2\right){}^2-2q^2 \left(3 m_{\alpha } m_{\beta }+m_{\alpha }^2+m_{\beta }^2\right)+(q^2)^2\right)-m_{\alpha} m_{\beta } m_{W'}^2 \big(\left(m_{\alpha }^2-m_{\beta }^2\right){}^2+2 q^2 \big(9m_{\alpha } m_{\beta }
\nonumber \\ &&
+2 m_{\alpha }^2+2 m_{\beta }^2\big)-5 (q^2)^2\big)+q^2 m_{\alpha}^2 m_{\beta }^2 \left(-4 m_{\alpha } m_{\beta }+m_{\alpha }^2+m_{\beta}^2-q^2\right)\big),
\end{eqnarray}
\begin{eqnarray}
f^j_{M,34}&=&m_{\alpha } \big(m_j^4 \left(-\left(m_{\alpha }-m_{\beta }\right) \left(m_{\alpha}+m_{\beta }\right){}^2 \left(m_{\alpha }+2 m_{\beta }\right)+q^2 \left(3 m_{\alpha }m_{\beta }+2 m_{\alpha }^2+5 m_{\beta }^2\right)-(q^2)^2\right)+m_j^2 \big((q^2)^2
\nonumber \\ &&
\times
\left(m_{\beta } \left(m_{\alpha }+3 m_{\beta }\right)-m_{W'}^2\right)+q^2 \left(\left(3m_{\alpha } m_{\beta }+2 m_{\alpha }^2+5 m_{\beta }^2\right) m_{W'}^2-m_{\beta }\left(5 m_{\alpha }^2 m_{\beta }-4 m_{\alpha } m_{\beta }^2+2 m_{\alpha }^3+m_{\beta}^3\right)\right)
\nonumber \\ &&
+\left(m_{\alpha }-m_{\beta }\right) \left(m_{\alpha }+m_{\beta}\right){}^2 \left(m_{\beta } \left(m_{\alpha } m_{\beta }+m_{\alpha }^2+2 m_{\beta}^2\right)-\left(m_{\alpha }+2 m_{\beta }\right) m_{W'}^2\right)\big)+(q^2)^2\big(m_{\alpha } m_{\beta }^3-m_{\beta } \big(3 m_{\alpha }
\nonumber \\ &&
+2 m_{\beta }\big)m_{W'}^2+2 m_{W'}^4\big)-q^2 \big(m_{\alpha } m_{\beta }^4 \left(m_{\beta }-3m_{\alpha }\right)+m_{\beta } \left(-13 m_{\alpha }^2 m_{\beta }-9 m_{\alpha } m_{\beta}^2-6 m_{\alpha }^3+2 m_{\beta }^3\right) m_{W'}^2
\nonumber \\ &&
+2 \left(3 m_{\alpha } m_{\beta }+2m_{\alpha }^2+5 m_{\beta }^2\right) m_{W'}^4\big)-\left(m_{\alpha }-m_{\beta }\right)\left(m_{\alpha }+m_{\beta }\right){}^2 \big(m_{\alpha }^2 m_{\beta }^3-2\left(m_{\alpha }+2 m_{\beta }\right) m_{W'}^4+m_{\beta } \big(m_{\alpha }
\nonumber \\ &&
+2 m_{\beta}\big) \left(3 m_{\alpha }+2 m_{\beta }\right) m_{W'}^2\big)\big),
\end{eqnarray}
\begin{eqnarray}
f^j_{M,5}&=&2 m_{\alpha } m_{\beta } \left(m_{\alpha }+m_{\beta }\right) \big(q^2 m_j^4 \left(m_{\alpha} m_{\beta }-4 m_{\alpha }^2-4 m_{\beta }^2+4 q^2\right)+m_j^2 \big(q^2 \big(-m_{\alpha}^3 m_{\beta }+m_{\alpha }^4+m_{\alpha }^2 \left(5 m_{\beta }^2-2 q^2\right)
\nonumber \\ &&
+m_{\alpha }m_{\beta } \left(q^2-m_{\beta }^2\right)+\left(m_{\beta }^2-q^2\right){}^2\big)+m_{W'}^2\left(\left(m_{\alpha }-m_{\beta }\right){}^2+3 q^2\right) \left(\left(m_{\alpha}+m_{\beta }\right){}^2-q^2\right)-9 q^2 m_{W'}^4\big)+3 q^2 m_j^6
\nonumber \\ &&
+q^2 m_{\alpha }^2 m_{\beta}^2 \left(m_{\alpha } m_{\beta }-m_{\alpha }^2-m_{\beta }^2+q^2\right)+m_{W'}^4 \left(2m_{\alpha }^4-4 m_{\alpha }^2 \left(m_{\beta }^2+q^2\right)-9 q^2 m_{\alpha } m_{\beta }+2\left(m_{\beta }^2-q^2\right){}^2\right)
\nonumber \\ &&
-m_{W'}^2 \left(\left(m_{\alpha }+m_{\beta}\right){}^2-q^2\right) \left(\left(3 m_{\alpha } m_{\beta }+2 m_{\alpha }^2+2 m_{\beta}^2\right) \left(m_{\alpha }-m_{\beta }\right){}^2+q^2 \left(m_{\alpha } m_{\beta }-2m_{\alpha }^2-2 m_{\beta }^2\right)\right)
\nonumber \\ &&
+6 q^2 m_{W'}^6\big),
\end{eqnarray}
\begin{equation}
f^j_{E,0}=m_{\alpha } m_{\beta } \left(m_{\alpha }-m_{\beta }\right) \left(m_{\alpha }^4-2 m_{\alpha }^2 \left(m_{\beta
   }^2+q^2\right)+\left(m_{\beta }^2-q^2\right){}^2\right) \left(m_j^2-m_{\alpha } m_{\beta }+2 m_{W'}^2\right).
\end{equation}
\begin{equation}
f^j_{E,12}=\left(m_{\alpha }-m_{\beta }\right) \left(m_j-m_{W'}\right) \left(m_j+m_{W'}\right) \left(m_{\alpha }^4-2 m_{\alpha }^2\left(m_{\beta }^2+q^2\right)+\left(m_{\beta }^2-q^2\right){}^2\right) \left(m_j^2-m_{\alpha } m_{\beta }+2m_{W'}^2\right),
\end{equation}
\begin{eqnarray}
f^j_{E,23}&=&\left(m_{\alpha }-m_{\beta }\right) \big((m_j^2 \big((q^2)^2 \left(4 m_{\alpha } m_{\beta }+m_{W'}^2\right)-q^2\big(m_{\alpha } m_{\beta } \left(2 m_{\alpha } m_{\beta }+3 m_{\alpha }^2+3 m_{\beta }^2\right)+2 \big(-3m_{\alpha } m_{\beta }+m_{\alpha }^2
\nonumber \\ &&
+m_{\beta }^2\big) m_{W'}^2\big)+\left(m_{\alpha }^2-m_{\beta }^2\right){}^2\left(m_{W'}^2-m_{\alpha } m_{\beta }\right)\big)+m_j^4 \left(m_{\alpha }^4-2 m_{\alpha }^2 \left(m_{\beta}^2+q^2\right)+6 q^2 m_{\alpha } m_{\beta }+\left(m_{\beta }^2-q^2\right){}^2\right)
\nonumber \\ &&
+m_{\alpha } m_{\beta } m_{W'}^2\left(\left(m_{\alpha }^2-m_{\beta }^2\right){}^2+2 q^2 \left(-9 m_{\alpha } m_{\beta }+2 m_{\alpha }^2+2 m_{\beta}^2\right)-5 (q^2)^2\right)+q^2 m_{\alpha }^2 m_{\beta }^2 \big(4 m_{\alpha } m_{\beta }+m_{\alpha }^2+m_{\beta}^2
\nonumber \\ &&
-q^2\big)-2 m_{W'}^4 \left(m_{\alpha }^4-2 m_{\alpha }^2 \left(m_{\beta }^2+q^2\right)+6 q^2 m_{\alpha } m_{\beta}+\left(m_{\beta }^2-q^2\right){}^2\right)\big),
\end{eqnarray}
\begin{eqnarray}
f^j_{E,34}&=&m_{\alpha } \big(m_j^4 \left(\left(m_{\alpha }-2 m_{\beta }\right) \left(m_{\alpha }-m_{\beta }\right){}^2\left(m_{\alpha }+m_{\beta }\right)+q^2 \left(3 m_{\alpha } m_{\beta }-2 m_{\alpha }^2-5 m_{\beta}^2\right)+(q^2)^2\right)+m_j^2 \big((q^2)^2 \big(m_{\beta } \big(m_{\alpha }
\nonumber \\ &&
-3 m_{\beta }\big)+m_{W'}^2\big)+q^2\left(m_{\beta } \left(5 m_{\alpha }^2 m_{\beta }+4 m_{\alpha } m_{\beta }^2-2 m_{\alpha }^3+m_{\beta}^3\right)+\left(3 m_{\alpha } m_{\beta }-2 m_{\alpha }^2-5 m_{\beta }^2\right) m_{W'}^2\right)+\big(m_{\alpha}
\nonumber \\ &&
-m_{\beta }\big){}^2 \left(m_{\alpha }+m_{\beta }\right) \left(m_{\beta } \left(-m_{\alpha } m_{\beta }+m_{\alpha}^2+2 m_{\beta }^2\right)+\left(m_{\alpha }-2 m_{\beta }\right) m_{W'}^2\right)\big)+(q^2)^2 \big(m_{\alpha }m_{\beta }^3+m_{\beta } \big(2 m_{\beta }
\nonumber \\ &&
-3 m_{\alpha }\big) m_{W'}^2-2 m_{W'}^4\big)+q^2 \big(-m_{\alpha }m_{\beta }^4 \left(3 m_{\alpha }+m_{\beta }\right)+m_{\beta } \left(-13 m_{\alpha }^2 m_{\beta }+9 m_{\alpha }m_{\beta }^2+6 m_{\alpha }^3+2 m_{\beta }^3\right) m_{W'}^2
\nonumber \\ &&
+2 \left(-3 m_{\alpha } m_{\beta }+2 m_{\alpha }^2+5m_{\beta }^2\right) m_{W'}^4\big)-\left(m_{\alpha }-m_{\beta }\right){}^2 \left(m_{\alpha }+m_{\beta }\right)\big(m_{\alpha }^2 m_{\beta }^3+2 \left(m_{\alpha }-2 m_{\beta }\right) m_{W'}^4
\nonumber \\ &&
+m_{\beta } \big(m_{\alpha }-2m_{\beta }\big) \left(3 m_{\alpha }-2 m_{\beta }\right) m_{W'}^2\big)\big),
\end{eqnarray}
\begin{eqnarray}
f^j_{E,5}&=&2 m_{\alpha } m_{\beta } \left(m_{\alpha }-m_{\beta }\right) \big(-q^2 m_j^4 \left(m_{\alpha } m_{\beta }+4 m_{\alpha}^2+4 m_{\beta }^2-4 q^2\right)+m_j^2 \big(q^2 \big(m_{\alpha }^3 m_{\beta }+m_{\alpha }^4+m_{\alpha }^2 \left(5m_{\beta }^2-2 q^2\right)+m_{\alpha }
\nonumber \\ &&
\times
\left(m_{\beta }^3-q^2 m_{\beta }\right)+\left(m_{\beta}^2-q^2\right){}^2\big)+m_{W'}^2 \left(\left(m_{\alpha }-m_{\beta }\right){}^2-q^2\right) \left(\left(m_{\alpha}+m_{\beta }\right){}^2+3 q^2\right)-9 q^2 m_{W'}^4\big)+3 q^2 m_j^6
\nonumber \\ &&
-q^2 m_{\alpha }^2 m_{\beta }^2 \left(m_{\alpha }m_{\beta }+m_{\alpha }^2+m_{\beta }^2-q^2\right)+m_{W'}^4 \left(2 m_{\alpha }^4-4 m_{\alpha }^2 \left(m_{\beta}^2+q^2\right)+9 q^2 m_{\alpha } m_{\beta }+2 \left(m_{\beta }^2-q^2\right){}^2\right)
\nonumber \\ &&
-m_{W'}^2 \left(\left(m_{\alpha}-m_{\beta }\right){}^2-q^2\right) \left(\left(m_{\alpha }+m_{\beta }\right){}^2 \left(-3 m_{\alpha } m_{\beta }+2m_{\alpha }^2+2 m_{\beta }^2\right)-q^2 \left(m_{\alpha } m_{\beta }+2 m_{\alpha }^2+2 m_{\beta }^2\right)\right)
\nonumber \\ &&
+6 q^2m_{W'}^6\big),
\end{eqnarray}

\section{Coefficients $g^{\alpha\beta}_k$ of the squared amplitude $|\overline{\cal M}_{\alpha\to\beta\sigma\sigma}|^2$}
\label{gApp}
In this Appendix, we provide the explicit expressions of the coefficients $g_k^{\alpha\beta}$ that are found in Eq.~(\ref{mnsqdamp}):
\begin{eqnarray}
g_1^{\alpha\beta}&=&-8 \big(2 m_{\alpha } m_{\beta } \left(2 m_{\sigma }^2+q^2\right)+m_{\alpha }^2 \left(2 m_{\beta}^2-q^2-2 (p_2+p_3)^2\right)-m_{\beta }^2 (q^2+2 (p_2+p_3)^2)+2 \big(m_{\sigma }^4
\nonumber \\ &&
-2 (p_2+p_3)^2 m_{\sigma }^2+(p_2+p_3)^2(q^2+(p_2+p_3)^2)\big)+(q^2)^2\big),
\end{eqnarray}
\begin{eqnarray}
g_2^{\alpha\beta}&=&-8 (q^2)^2 \big(-2 m_{\alpha } m_{\beta } \left(2 m_{\sigma }^2+q^2\right)+m_{\alpha }^2 \left(2m_{\beta }^2-q^2-2 (p_2+p_3)^2\right)-m_{\beta }^2 (q^2+2 (p_2+p_3)^2)+2 \big(m_{\sigma }^4
\nonumber \\ &&
-2 (p_2+p_3)^2 m_{\sigma }^2+(p_2+p_3)^2(q^2+(p_2+p_3)^2)\big)+(q^2)^2\big),
\end{eqnarray}
\begin{eqnarray}
g_3^{\alpha\beta}&=&8 \big(-2 q^2 m_{\alpha } m_{\beta } \left(2 m_{\sigma }^2+q^2\right)+m_{\alpha }^4 \left(2 m_{\sigma}^2+q^2\right)+m_{\beta }^4 \left(2 m_{\sigma }^2+q^2\right)-m_{\alpha }^2 \left(4 m_{\beta }^2m_{\sigma }^2+q^2 (q^2+2 (p_2+p_3)^2)\right)
\nonumber \\ &&
-q^2 m_{\beta }^2 (q^2+2 (p_2+p_3)^2)-2 q^2 \left(m_{\sigma }^2-(p_2+p_3)^2\right)\left(-m_{\sigma }^2+q^2+(p_2+p_3)^2\right)\big),
\end{eqnarray}
\begin{eqnarray}
g_4^{\alpha\beta}&=&8 \big(2 q^2 m_{\alpha } m_{\beta } \left(2 m_{\sigma }^2+q^2\right)+m_{\alpha }^4 \left(2 m_{\sigma}^2+q^2\right)+m_{\beta }^4 \left(2 m_{\sigma }^2+q^2\right)-m_{\alpha }^2 \left(4 m_{\beta }^2m_{\sigma }^2+q^2 (q^2+2 (p_2+p_3)^2)\right)
\nonumber \\ &&
-q^2 m_{\beta }^2 (q^2+2 (p_2+p_3)^2)-2 q^2 \left(m_{\sigma }^2-(p_2+p_3)^2\right)\left(-m_{\sigma }^2+q^2+(p_2+p_3)^2\right)\big),
\end{eqnarray}
\begin{equation}
g_5^{\alpha\beta}=-16 \left(m_{\alpha }+m_{\beta }\right) \left(2 m_{\sigma }^2+q^2\right) \left(\left(m_{\alpha}-m_{\beta }\right){}^2-q^2\right),
\end{equation}
\begin{equation}
g_6^{\alpha\beta}=16 q^2 \left(m_{\alpha }-m_{\beta }\right) \left(2 m_{\sigma }^2+q^2\right) \left(\left(m_{\alpha}+m_{\beta }\right){}^2-q^2\right).
\end{equation}

\end{document}